# Model ensembles of artificial neural networks and support vector regression for improved accuracy in the prediction of vegetation conditions

**Chrisgone Adede [1,*], Robert Oboko [1], Peter W. Wagacha [1] and Clement Atzberger [2]**

[1] School of Computing and Informatics, University of Nairobi (UoN), P.O. Box 30197, GPO, Nairobi, Kenya; roboko@uonbi.ac.ke (R.O.); waiganjo@uonbi.ac.ke (P.W.W.)
[2] National Drought Management Authority (NDMA), Lonrho House -Standard Street, Box 53547, Nairobi 00200, Kenya
[3] Institute of Surveying, Remote Sensing and Land Information, University of Natural Resources (BOKU), Peter Jordan Strasse 82, A-1190 Vienna, Austria; clement.atzberger@boku.ac.at
* Correspondence: adedekris@gmail.com

**Abstract:** There is increasing need for highly predictive and stable models for the prediction of drought as an aid to better planning for drought response. This paper presents the performance of both homogenous and heterogenous model ensembles in the prediction of drought severity using the study case techniques of artificial neural networks (ANN) and support vector regression (SVR). For each of the homogenous and heterogenous model ensembles, the study investigates the performance of three model ensembling approaches: linear averaging (non-weighted), ranked weighted averaging and model stacking using artificial neural networks. Using the approach of "over-produce then select", the study used 17 years of data on 16 selected variables for predictive drought monitoring to build 244 individual ANN and SVR models from which 111 models were selected for the building of the model ensembles. The results indicate marginal superiority of heterogenous to homogenous model ensembles. Model stacking is shown to realize models that are superior in performance in the prediction of future vegetation conditions as compared to the linear averaging and weighted averaging approaches. The best performance from the heterogenous stacked model ensembles recorded an $R^2$ of 0.94 in the prediction of future vegetation conditions as compared to an $R^2$ of 0.83 and $R^2$ of 0.78 for both ANN and SVR respectively in the traditional champion model approaches to the realization of predictive models. We conclude that despite the computational resource intensiveness of the model ensembling approach to drought prediction, the returns in terms of model performance is worth the investment, especially in the context of the recent exponential increase in computational power.

## 1. Introduction

One point of divergence in the practice of drought monitoring is the definition of drought. The differences in the definition of drought is informed by the variance in interest on and the attendant types of drought that is any of meteorological, hydrological, agricultural or socio-economic as documented in UNOOSA (2015). Despite the lack of a standard definition, the need for drought monitoring is well understood in the context of loses arising from their occurrence and the need for planned action. Losses from past droughts are documented, for example in Government of Kenya (2012) and Cody (2010) with a detailed review of a range of impacts in Ding, Hayes & Widhalm (2011).

Drought monitoring happens in the context of drought early warning systems (DEWS) that are increasingly either near real time (NRT) or ex-ante (predictive). NRT systems, for example include the univariate BOKU system (Klisch & Atzberger, 2016) and the Famine Early Warning Systems Network (FEWSNET) (Brown et al, 2015) that are based on MODIS vegetation datasets and the US drought monitor in Svoboda et al (2002) which is a multi-variate system.

Predictive systems are either based on a single variable/index, a multi-variable index or on multiple indices (variables). Single variable indexes, especially that use the standardised precipitation index (SPI) are deployed in Ali et al. (2017) and Khadr (2016). The single variable indices are generally easy to interpreted as comparted to multi-variable (super-index) indices like the case of the Enhanced Combined Drought Index (ECDI) (Enenkel, 2016) that integrates four input datasets (rainfall, soil moisture, land surface temperature and vegetation status). Increasingly, the use of multiple variables in predictive drought system is gaining currency in an approach where multiple indices are used in to build the predictive systems. Such systems include the us of 11 variables and indices in Tadesse et al. (2014) and that in Adede et al. (2019) that uses 10 variables in the prediciton of future vegetation conditions.

The increasing popularity of predictive DEWs in the light of increased damages suffered from droughts together with the proliferation of multiple indices and variables for drought monitoring has led to the need for multi-variate models that are both highly predictive and stable over time to support proactive drought risks management (DRM) initiatives. One way to improve stability and accuracy of models is model ensembling. Variously, model ensembling is defined as the formulation of multiple individually trained models and the subsequent combination of their outputs (Cortes, Kuznetsov & Mohri, 2014; Dietterich, 2000; Opitz & Maclin, 1999 and Re & Valentini, 2012). In this sense, model ensembling is akin to the innate human nature of seeking multiple opinions in decision making. Model ensembling, therefore, aims to produce more accurate and more stable predictors that arise from the ability to average out noise and therefore achieve better generalizability (Güneş, 2017; Opitz & Maclin, 1999).

The common issues in model ensembling include the process for over-production of ensemble base models, the selection of the based model for ensemble membership and the combination of the outputs of the ensemble members. The process of over-production of models has hyper-parameter tuning at the core. Hyper-parameter tuning becomes even more critical in the instances when automation of model building and selection is a key objective like in Nay, Burchfield & Gilligan (2018) in which three (3) hyper-parameters needed to be tuned for gradient boosted machine (GBM). The problem of the selection of ensemble membership deals with the question of which sub-set of the models from the model over-production process offers the best predictive power. The problem of ensemble membership selection is for example documented in Partalas, Tsoumakas & Vlahavas (2012), Re & Valentini (2012) and Reid (2007). The distinct approach to ensemble member prediction includes greedy search (Partalas, Tsoumakas & Vlahavas, 2012), that realizes the global best sub-set by taking local optimal decision when changing the current set, ensemble pruning in Re & Valentini (2012) that uses both statistical and semi-definite programming approaches and the statistical ensemble method in Escolano et al. (2003) that uses resampling to estimate accuracy of individual members and multiple comparisons to choose ensemble membership. Generally, it is agreed that no single approach fits all when it comes to ensemble member selection.

Related work on ensemble modeling includes the use of bagging and boosting (Belayneh et al. ,2016) and Opitz & Maclin (1999). While both bagging and boosting are ensembles that aim to generate strong learners from weak learners, the major difference is that bagging averages predictions from multiple sub-sets of the training data with the aim to reduce variance while boosting sequentially learns predictors, first from the whole data set then subsequently on the training set based on previous performance with the aim to increasing bias. Apart from bugging and boosting, an increasingly common approach to model ensembling is stacking that uses a meta/ super leaner to combine weak base predictors to reduce generalization error. Dzeroski & Zenko (2004) documents stacking as having better performance as compared to the selection of the the best classifier. The study in Belayneh et al. (2016), for example, uses both bagging and boosting in drought prediction using wavelet transforms while Ganguli & Reddy (2014) used the copula method on support vector regression (SVR) to simulate ensembles of drought forecasts. Common to both Belayneh et al. (2016) and Ganguli & Reddy (2014) is the use of a single drought index in the prediction of meteorological drought. On the contrary, the systems in Wardlow et al (2012) and Tadesse et al. (2010), for example, use multiple indices in the forecasting of future vegetation conditions.

Model ensembles can either be homogenous thus based on same technique (Adhikari & Agrawal, 2013) or heterogenous and hence multi-technique (Reid, 2007). The objective of this study is to investigate the performance of both homogenous and heterogenous multi-variate model ensembles in the prediction of vegetation conditions 1 month ahead as a proxy to drought conditions using ANN and SVR as the chosen case study techniques. The homogenous and heterogenous model ensembles are realized using three different ensemble approaches: non-weighted averaging, weighted averaging and ANN driven model stacking.

## 2. Material and Methods

### *2.1 Study Area*

The study area (Figure 1) is also documented in Adede et al. (2019) as covering an area of over 215000km² and is located in Northern Kenya and covers four counties that are classified as arid counties within Kenya's arid and semi-arid lands (ASALs). The area is characterized by both low rainfall and low vegetation cover with an average normalized difference vegetation index (NDVI) of below 0.4 except for May and Nov that peaks at 0.43. The rainfall averages around 250mm for Turkana, Marsabit and Mandera and a little higher at around 370mm in Wajir. The region experiences a bimodal rainfall pattern with the two seasons in March-May (MAM) and October-December (OND). Across the counties in the region, 5-6 months are considered wet.

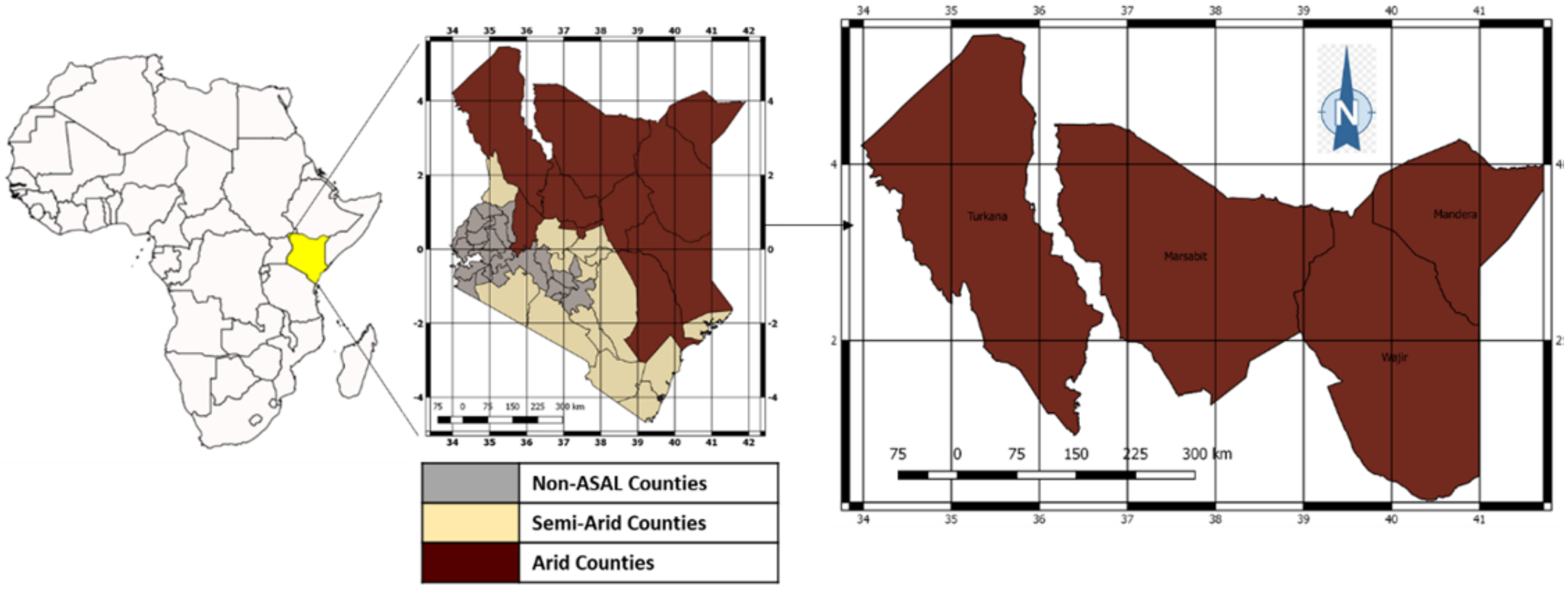

Figure 1: The study area (to the right) and its location within Kenya. The inset (left) provides the location of Kenya in Africa while the map of Kenya (center) shows the grouping of the Kenyan counties into arid and semi-arid lands (ASAL) and non-ASAL.

### *2.2 Pre-modelling*

#### 2.2.1 The Data

The study uses 22 variables and indices derived from precipitation, vegetation, temperature and evapotranspiration datasets. The variables are grouped into three categories: vegetation, precipitation and water balance (influencer) datasets. The data used in the study covered the period March 2001 to December 2017 and are as provided in Table 1.

Table 1: The study base datasets- categories, sources and description.

| Category | Base dataset | Source | Description |
|---|---|---|---|
| Vegetation | Normalised Difference Vegetation Index (NDVI). | Land Processes Distributed Active Archive Center (LPDAAC) as documented in Didan (2015a) and Didan (2015b) | Combination of both MODIS Terra (MOD13Q1) & MODIS Aqua (MCD13Q1) using the Whittaker smoothing approach (Klisch & Atzberger, 2016) |
| Precipitation | Rainfall Estimates (RFE) | RFE from both Tropical Applications of Meteorology using SATellite (TAMSAT) and Climate Hazard Group InfraRed Precipitation (CHIRPS) as documented in Tarnavsky et al. (2014) and Funk et al. (2014) respectively. | TAMSAT version 3.0 product & CHIRPS version 2.0 product aggregated and spatially sub-set as in the BKO system. |
| Water Balance | Land Surface Temperature data (LST) | LPDAAC (Wan, Hook & Hulley, 2015) | MODIS Terra Land Surface Temperature/Emissivity 8-Day L3 Global 1km SIN Grid V006 product (MOD11A2) |
| | Evapotranspiration (EVT) | LPDAAC (Running, S., Mu, Q., Zhao, M., 2017) | MODIS/Terra Net Evapotranspiration 8-Day L4 Global 500m SIN Grid V006 (MOD16A2) |
| | Potential Evapotranspiration (PET) | LPDAAC (Running, S., Mu, Q., Zhao, M., 2017) | MODIS/Terra Net Evapotranspiration 8-Day L4 Global 500m SIN Grid V006 (MOD16A2) |
| | Standardized Precipitation-Evapotranspiration (SPEI) | Be SPEI Global Drought Monitor (Beguería et al ,2014) | The difference between precipitation and potential evapotranspiration standardised and used like in in standardised precipitation index (SPI) |

2.2.2 The variables

The above datasets are transformed to indices/ variables that are then used for the predictive study following two approaches- the calculation of relative range difference approach and the standardization approach. The relative range difference is calculated as shown in equation 1. The standardization approach, on the other hand, involves fitting the bases dataset to an appropriate probability distribution so that the mean of the transformed variable is zero (0) and the standard deviation is equivalent to (1). Both the relative range and standardised transformations are done at pixel level ($c$) for each time step ($i$) prior to aggregation.

$$RR_h(c,i) = 100 * (\frac{[X(c,i) - MIN(c,i)]}{[MAX(c,i) - MIN(c,i)]}) \quad (1)$$

*Where $RR_h$ is the scaled relative range difference, X(c,i) the current value, MIN (c,i) the historical minimum and MAX (c,i) the historical maximum.*

The variables/ indices used in this predictive modelling study are from the datasets above and are described in Table 2.

Table 2: Variables used in the study to predict vegetation condition index. Near infrared (NIR) and Red are the spectral reflectances in near infrared and red spectral channels of MODIS satellite.

| No | Variable | Variable Description | Index Calculation |
|---|---|---|---|
| 1 | NDVIDekad | *NDVI for last dekad of the month* | *NDVI = (NIR-Red)/(NIR+Red)* |
| 2 | VCIdekad | *VCI for the last dekad of the month* | *Transformation of NDVI following on Equation (1)* |
| 3 | VCI1M | *VCI aggregated over the month* | *Transformation of NDVI following on Equation (1)* |
| 4 | VCI3M | *VCI aggregated over the last 3 months* | *Transformation of NDVI following on Equation (1)* |
| 5 | TAMSAT_RFE1M | *TAMSAT Rainfall Estimate aggregated over the month* | *RFE from TAMSAT version 3 product (in mm) Tarnavsky et al. (2014)* |
| 6 | TAMSAT_RFE3M | *TAMSAT Rainfall Estimate aggregated over the last 3 months* | *RFE from TAMSAT version 3 product (in mm) Tarnavsky et al. (2014)* |
| 7 | TAMSAT_RCI1M | *TAMSAT Rainfall Condition Index aggregated over the last 3 months* | *TAMSAT RFE calculated using Equation (1)* |
| 8 | TAMSAT_RCI3M | *TAMSAT Rainfall Condition Index aggregated over the last 3 months* | *TAMSAT RFE calculated using Equation (1)* |
| 9 | TAMSAT_SPI1M | *TAMSAT Standardized Precipitation Index aggregated over the last 1 month* | *TAMSAT RFE transformed to a normal distribution so that SPImean c,i = 0. WMO (2012)* |
| 10 | TAMSAT_SPI3M | *TAMSAT Standardized Precipitation Index aggregated over the last 3 months* | *TAMSAT RFE transformed to a normal distribution so that SPImean c,i = 0. WMO (2012)* |
| 11 | CHIRPS_RFE1M | *CHIRPS Rainfall Estimate aggregated over the month* | *RFE from CHIRPS version 3 product (in mm) Funk et al. (2014)* |
| 12 | CHIRPS_RFE3M | *CHIRPS Rainfall Estimate aggregated over the last 3 months* | *RFE from CHIRPS version 3 product (in mm) Funk et al. (2014)* |
| 13 | CHIRPS_RCI1M | *CHIRPS Rainfall Condition Index aggregated over the last 1 month* | *CHIRPS RFE calculated using Equation (1)* |
| 14 | CHIRPS_RCI3M | *CHIRPS Rainfall Condition Index aggregated over the last 3 months* | *CHIRPS RFE calculated using Equation (1)* |
| 15 | CHIRPS_SPI1M | *CHIRPS Standardized Precipitation Index aggregated over the last 1 month* | *CHIRPS RFE transformed to a normal distribution so that SPImean c,i = 0. WMO (2012)* |
| 16 | CHIRPS_SPI3M | *CHIRPS Standardized Precipitation Index aggregated over the last 3 months* | *Same as Index No. 15* |
| 17 | LST1M | *Land Surface Temperature aggregated over the month* | *Average LST over the last one month* |
| 18 | EVT1M | *Evapotranspiration aggregated over the month* | *Average MODIS EVT over the last one month* |
| 19 | PET1M | *Potential Evapotranspiration aggregated over the month* | *Average MODIS PET over the last one month* |
| 20 | TCI1M | *Temperature Condition Index aggregated over the month* | *MODIS LST transformed using Equation (1)* |
| 21 | SPEI1M | *Standardized Precipitation Evapotranspiration aggregated over the month* | *Follows the standardization approach on the difference between precipitation*$(P_i)$ *and potential evapotranspiration* $(PET_i)$ *using the log logistic probability distribution* |
| 22 | SPEI3M | *Standardized Precipitation Evapotranspiration aggregated over the last 3 months* | *Follows the standardization approach on the difference between precipitation*$(P_i)$ *and potential evapotranspiration* $(PET_i)$ *using the log logistic probability distribution* |

Variables 1-4 are vegetation indices while variables 5-16 to variables are two sets of precipitation indices from TAMSAT (5-10) and CHIRPS (11-16) respectively. The study methodology is designed to select between TAMSAT and CHIRPS for the modeling process. Variables 17-22 are the commonly used together with the vegetation and precipitation indices in predictive drought modeling and are used in this study as influencer variables.

The vegetation datasets in Table 1 (variable 1-4) are smoothed using a modified Whittaker smoothing algorithm and are directly sourced from the Institute for Surveying, Remote Sensing and Land Information, University of Natural Resources and Life Sciences (BOKU). The vegetation datasets are calculated at pixel level prior to aggregation in both time and space. The vegetation condition index (VCI) variables are calculated following on the relative range formula in Equation 1. The precipitation datasets from both CHIRPS and TAMSAT are also calculated at pixel level prior to aggregation. While the RCI based datasets (variables 7,8,13 & 14) follow on the Equation 1, the SPI variables (9,10,15 & 16) have the base dataset (RFE) fitted to a probability distribution then transformed to a normal distribution so that the SPI has a mean of zero (0) and a standard deviation of one (1) as recommended by WMO( 2012). The SPEI datasets are also standardised following the log logistic probability distribution (Beguería et al., 2014; Vicente-Serrano et al., 2010).

2.2.3 The learning scenario

To be able to develop predictive drought monitoring models, the phenomenon of drought needs to well defined. From the literature review, drought has key characteristics in its definition including spatial coverage, a severity dimension and temporal aspect. With these key concepts, we formulate the predictive drought monitoring problem using Equation 2.

$$D_{(i,j)} = f(x_1, x_2, x_3 \cdots x_n) \quad (2)$$

Where $D_{(i,j)}$ is a quantification of drought severity (intensity) for a spatial extent *i* at time *j*, $f$ is a function that accepts a set of n (n ≥1) variables and transforms them to approximate the real value for drought severity $D_{(i,j)}$. The $n$ variables $x_1, x_2, x_3 \cdots x_n$ are predictor variables that are used in drought monitoring.

The learning scenario in this study therefore involves the need to define $D_{(i,j)}$ for all the four counties in the study area and to over-produce and select the appropriate $f's$ to be selected for model ensembling. We interpret f as machine learning (ML) technique generated functions using artificial neural networks (ANN) and support vector regression (SVR) as case study techniques.

2.2.4 Definition of drought ($D_{(\mathrm{i},j)}$)

Given that drought severity, $D_{(i,j)}$), cannot be directly quantified, the practice is to use proxy index(es) in its quantification. The most common proxies used in the definition of drought ($D_{(i,j)}$) are the standardised precipitation index (SPI) and the Vegetation Condition Index (VCI), both sets of which data and variables are part of the study. Based on McKee et al. (1993), the SPI standardizes precipitation through an initial transformation and subsequent fitting to a normal distribution. The results are values typically between -3 and +3 with higher values indicating wet conditions. The VCI (Liu & Kogan, 1996; Klisch & Atzberger, 2016), on the other hand, is a relative range normalization of the NDVI in the 0–1 range (both end points included) that is generally scaled to the 0-100 range. The occurrence of a new minimum or maximum will, however, result in values below 0 and beyond 100 respectively.

The study considered the SPI and the VCI approach in the definition of drought on the 3-month aggregation of the indices. From the properties of both the SPI and VCI described above, we chose the VCI as the basis of definition of drought for three reasons. First, is the ease in its interpretation, that with a range of 0 to 100. Second is the fact that, as an index, it is used in the monitoring of agricultural drought which is a later stage drought as compared to meteorological drought indicated by SPI and; third is the fact that it is a measured quantity as opposed to the SPI that is, for the case of this study, is a modelled quantity. The task of drought prediction is therefore expressed as the task of predicting future (1 month ahead) VCI values aggregated over three months (VCI3M) using lagged values of the study variables.

*2.3 Methods*

2.3.1 Variable selection

The study processed a duplicate set of precipitation variables, both from TAMSAT and CHIRPS. The selection of variables in this study was thus formulated as the selection between TAMSAT and CHIRPS variables for the prediction of future drought intensity as defined by VCI3M.

Multiple methods were used in the selection between the two datasets to quantify the relationship between the variables and vegetation conditions as quantified by VCI3M. Variable selection was, however, preceded by the establishment of appropriateness for purpose of the methods through the test for normality using both density plots and Shapiro Wilk test. Subsequently, correlational analysis, step-wise regression, Akaike information criterion (AIC), Relative importance (RImp) of variables and a modelling approach to variable selection were used to establish which datasets between TAMSAT and CHIRPS to use in the modelling process.

2.3.2 Modelling methodology

With the drought prediction problem formulated as in Equation 2, the modelling methodology is reduced to the search for all the $f's$ that approximate the drought severity ($D_{(i,j)}$). This definition mirrors that of the generic machine learning (ML) problem and, in the context of model ensembling using the over-produce, select and combine approach, it equates a model space search for all the functions ($f's$) from the set of all functions ($F_U$) deemed to approximate ($D_{(i,j)}$) with some degree of accuracy measured by some model performance metrics. The functions ($f's$), are from the family of artificial neural networks (ANN) and support vector regression (SVR), in the context of this study.

The above formulation gives rise to the following key concepts. The *variable space* $(x_1, x_2, x_3 \cdots x_n)$ corresponds to all the variables used to measure ($D_{(i,j)}$), *the model space ($F_U$)* is the set of all possible models that can be derived from all the possible combination of the variables deemed to measure both ($D_{(i,j)}$) using the machine learning technique M that is either ANN or SVR.

*2.3.3 Model building*

Bagging, as a method of model ensembling is used as a standard to build both the ANN and SVR models in a setup in which the training dataset is resampled as indicated in the model building process presented in Figure 2. Not indicated in the model building process is normalization of the variables prior to building of the models to ensure the input variables were all at a comparable range. The main steps in the models building process (Figure 2) are- sampling, actual model building and model performance evaluation for both ANN and SVR techniques.

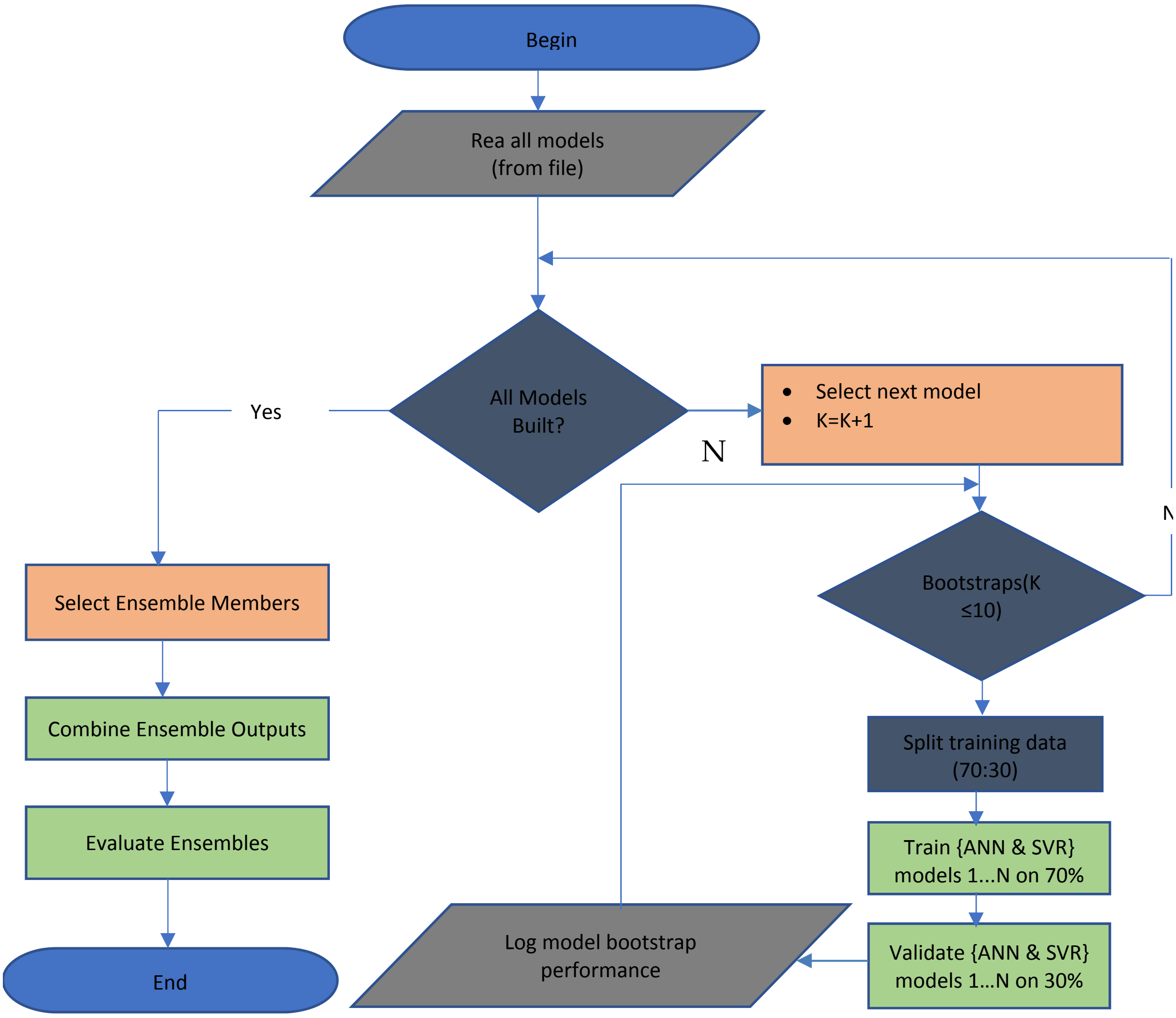

Figure 2: Model building process from model over-production to model selection for ensemble membership. The actual model building process using both ANN and SVR are preceded by a model space reduction process undertaken by two distinct steps. First, the formulation of assumptions and second, the use of a cut-off criteria on models considered predictive enough to be include in the ensemble.

2.3.3.1 Model space reduction & the ensemble membership

Post selection between TAMSAT and CHIRPS variables, the study has 16 variables. The initial cardinality of the modelling space, with VCI3M as the target variable, is a massive 65,535 models from all the 16 lagged predictors. The distribution of the models by count for each given number of variables, from 1 to 16 is as illustrated in Figure 3.

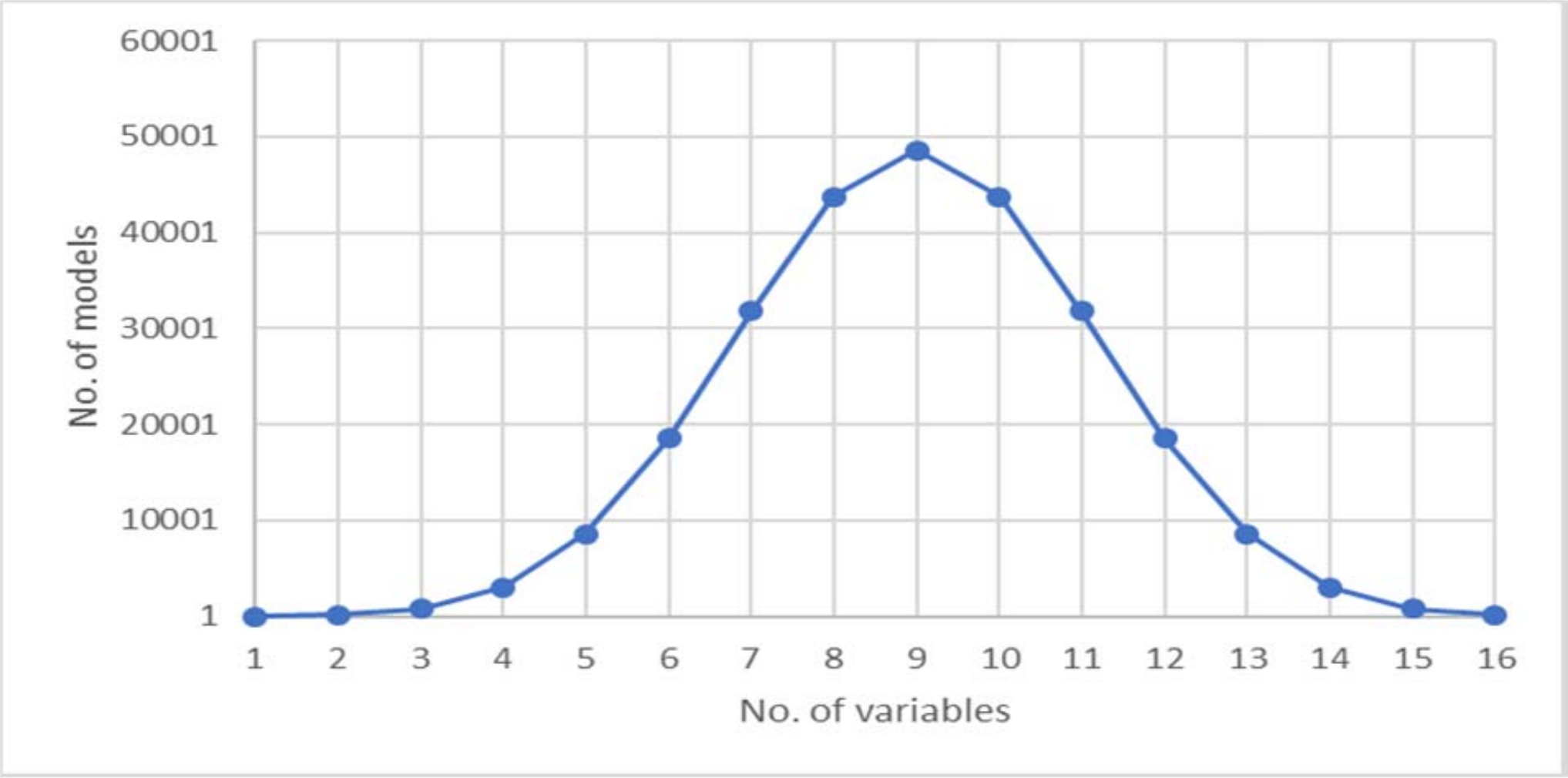

Figure 3: Model space of the drought severity prediction problem. The figure indicates the number of models of each length for a total of 65,535 models.

To achieve a reduction in the cardinality of the model space, we group the variables into categories and follow this up with assumptions. The variables are grouped into three categories: precipitation (VCI3M, NDVIDekad, VCI1M & VCIdekad), vegetation (RFE1M, RFE3M, RCI1M, RCI3M, SPI1M & SPI3M) and influencer variables (LST1M, EVT1M, PET1M, TCI1M, SPEI1M & SPEI3M). The influencer variables have an element of either accounting for impact of temperature on drought severity or indicating water balance as a function of both supply and demand.

The key assumption made by the study is that for clarity and interpretability of models, one variable of same category is sufficient in a single model. This assumption was validated for soundness in Adede et al. (2019). The process of model space reduction is presented in Figure 4.

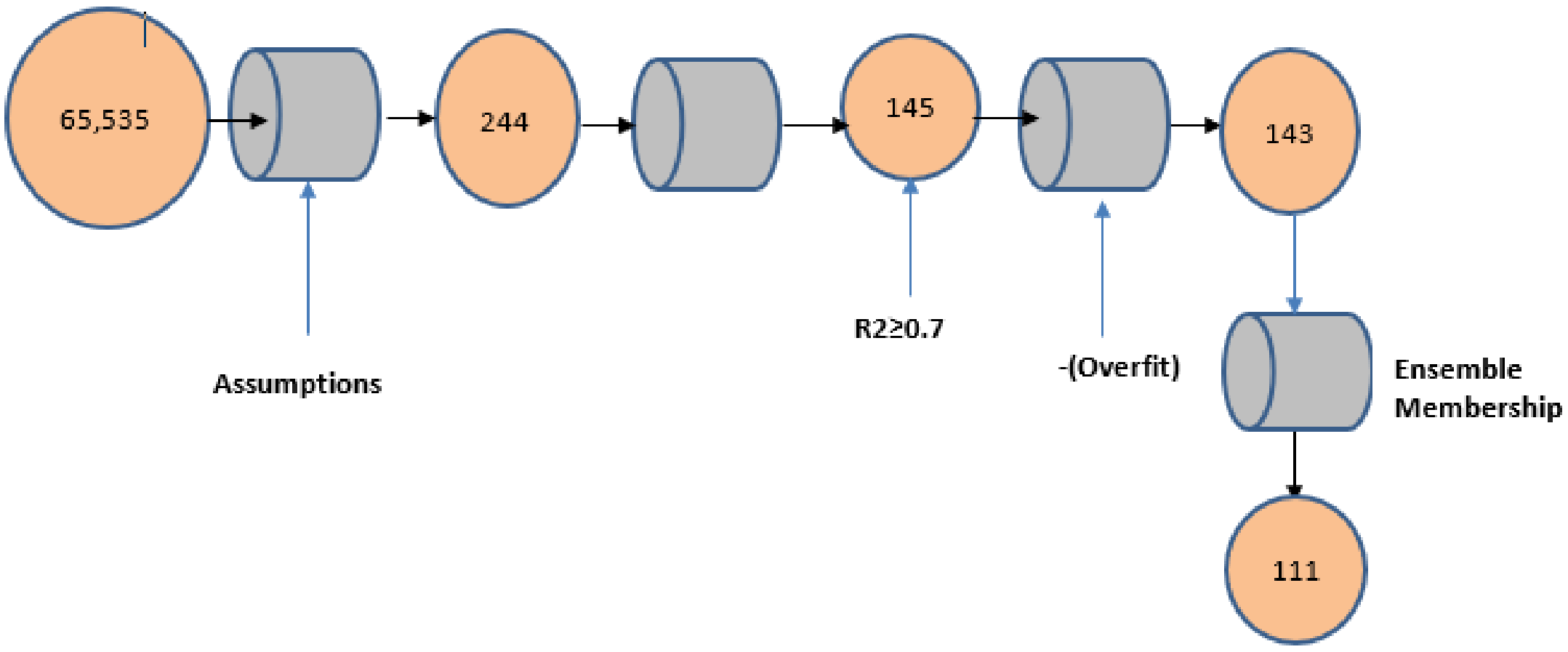

Figure 4: The model space reduction process. The assumption of including only one variable of a type: precipitation, vegetation and those of water balance (influencer variables) massively reduces the set of possible models from 65,535 to 244.

All the 244 models were build using both the ANN and SVR techniques A further reduction was achieved by only retaining the 145 models that had $R^2 \geq 0.7$ from both the ANN and SVR techniques. The final set was reduced by two models, to 143, from the ANN technique that we judged to have more than a 3% loss in performance between training and validation and were thus overfit.

Given that multi-model ensembles combine the predictions from different models and the fact there is need to search for the smallest sub-set that is most predictive, we used the 143 models from the ANN process as the basis for the selection of the ensemble membership. The ensemble membership was realized from an experimental process. The experimental process involved the ranking of all the models by descending $R^2$, the iterative elimination of the lowest ranked models in batches of 5 from the ensemble and a recalculation of the performance of the ensemble as measured by $R^2$. The elimination stops when a reduction in performance is realized. After the recording of a drop-in performance, the last batch of 5 are eliminated and added to the model one at a time from the best former only for that that do not degraded ensemble performance. The membership is then chosen as the least membership for which $R^2$ is greedily maximized.

*2.3.3.2 Sampling process*

To build the requisite set of models, the dataset (March 2001 to December -2017) is partitioned into in-sample and out-sample data in an approach similar to that used in Adede et al. (2019) and Nay, Burchfield & Gilligan (2018). The in-sample data is, subsequently, repeatedly and randomly split into training and validation datasets in the ration of 70:30 for each iteration of the model building process. The partitioning of the dataset is as shown in Figure 5.

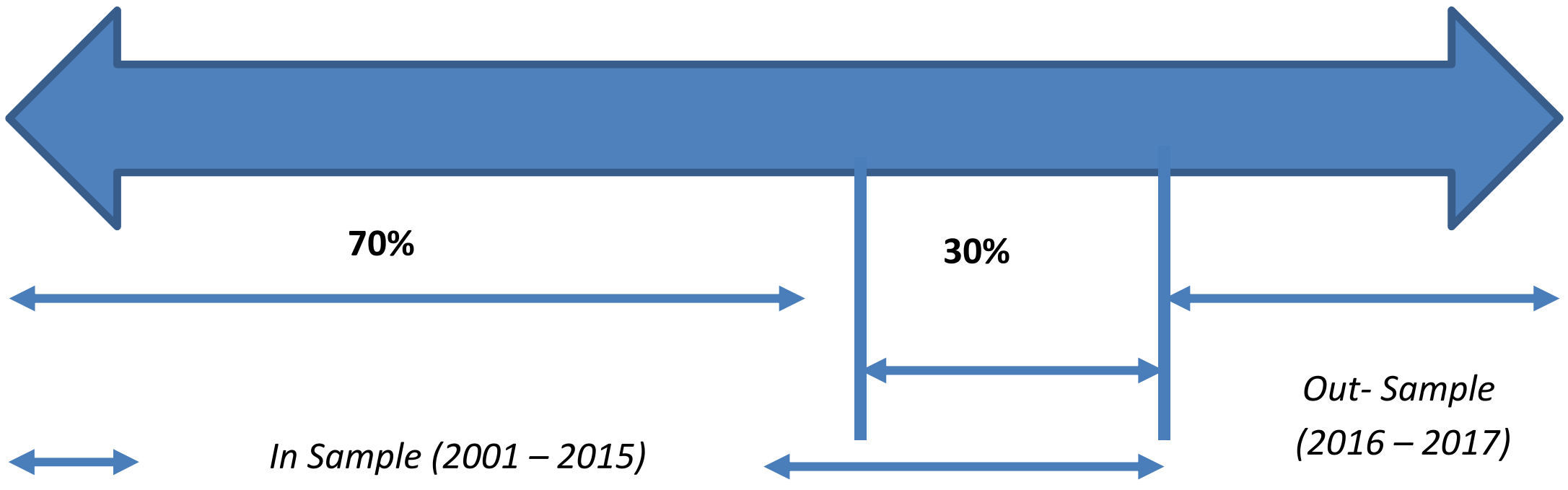

Figure 5: The in-sample (70:30) approach to model building and validation with a separate (out-sample) dataset for model testing.

*2.3.3.3 ANN and SVR model building process- model parameters and model assessment*

The models formulated after all assumptions were accounted for were subjected to an automated brute-force process that built, assessed and logged the performance of each of the models, both for the ANN and for the SVR case study techniques.

The choice of the model hyper-parameters for the ANN process followed the rule of thumb in Huang (2003) together with an experimental process in obtaining the appropriate number of hidden layers. The ANN were trained using resilient backpropagation (RPROP). RPROP is documented to have faster convergence speed, accuracy and robustness without the requirement for parameter tuning as documented Riedmiller & Braun (1993) and in Chen & Lin (2011). For SVR, we did an initial run that realized the best performance of all the models at an epsilon of 0.2 and a cost parameter of 32. The search for the single optimal configuration for the SVR technique followed the grid search approach using the statistical computing software- R.

Model performance evaluation was done using the determinant of correlation ($R^2$). The error measures like mean absolute error (MAE), root mean squared error (RMSE), mean absolute percentage error (MAPE) were also calculated.

### 2.3.4 Model ensembling approaches

The different models built were selected and their scores combined in two distinct approaches: homogenous ensembles in which only models from one technique are combined and heterogenous ensembles that combines the outputs from both ANN and SVR techniques.

For both the homogenous and heterogenous ensembles, three methods of ensembling were investigated including linear averaging (non-weighted), rank weighted averaging and model stacking. While linear averaging assumes similar weights for the individual models, weighted averaging uses the performance of the individual models in evaluation to assign weights on their prediction on the out-of-sample data. The linear average approach and the rank weighted approach follow the equations in 3 and 4 respectively.

$$\frac{1}{n}\sum_{i=1}^{n} p_i \text{ ............ } (3)$$

where $p_i$ as the prediction from the $i^{th}$ model and $n$ is the number of models in the ensemble

$$\frac{1}{n}\sum_{i=1}^{n} p_{i*} w_i \text{............ } (4)$$

Where $w_i$ is the normalized weight for each model in the ensemble such that $\sum_{i=1}^{n} w_i = 1$. The weights are therefore stretched between 0 and 1 centred around min with the max as the scale.

The model stacking approach, however, builds a meta-model using the ANN process to learn weights for the individual models using a perceptron learning approach.

### 3.3.6 Performance evaluation of the model ensembles

The process for evaluating the performance of the model ensembles had the following steps- (1) the selection of a common measure of performance; (2) the identification of the base models for comparison and; (3) scenarios of performance of the best individual models as compared to model ensembles. Performance in regression was evaluated using $R^2$ while performance in classification was evaluated using both accuracy and area under the receiver operating characteristic curve (AUROC). The best models from both the ANN and SVR processes were used as the base models. Performance in classification was analyzed following the five (5) vegetation deficit classes defined on VCI3M as used in Klisch, Atzberger & Luminari (2015), Klisch & Atzberger (2016), Meroni et al (2019) and Adede et al. (2019). This classification is presented in Table 3.

Table 3: Drought classes using the approach in Klisch, Atzberger & Luminari (2015), Klisch & Atzberger (2016), Meroni et al (2019) and Adede et al. (2019).

| **VCI3M** Limit Lower | **VCI3M** Limit Upper | **Description of Class** | **Drought Class** |
|---|---|---|---|
| ≤0 | <10 | Extreme vegetation deficit | 1 |
| 10 | <20 | Severe vegetation deficit | 2 |
| 20 | <35 | Moderate vegetation deficit | 3 |
| 35 | <50 | Normal vegetation conditions | 4 |
| 50 | ≥100 | Above normal vegetation conditions | 5 |

## 3. Results and Discussion

### *3.1 Selection between TAMSAT & CHIRPS*

The choice between TAMSAT and CHIRPS was undertaken using multiple methods. While the SPI variables are normalized, the other variables were tested to normality using the Shapiro Wilk test for normality. The Shapiro-Wilk test results are as provide in Table 4.

Table 4: Shapiro-Wilk test on normalized CHIRPS and TAMSAT datasets

| No | Variable | p-value |
|---|---|---|
| 1 | TAMSAT_RFE1M | 0.0000 |
| 2 | CHIRPS_RFE1M | 0.0000 |
| 3 | TAMSAT_RFE3M | 0.0000 |
| 4 | CHIRPS_RFE3M | 0.0000 |
| 5 | TAMSAT_RCI1M | 0.0000 |
| 6 | CHIRPS_RCI1M | 0.0000 |
| 7 | TAMSAT_RCI3M | 0.0000 |
| 8 | CHIRPS_RCI3M | 0.0000 |

With the null hypothesis corresponding to the sample drown belonging to a normally distributed population, we reject normality for all cases where the p-value≥0.05 at $\alpha$=0.05. The non-SPI variables are not normally distributed. Given the mix of normally distributed and non-normally distributed variables, the choice of methods for analysis therefore used non-parametric methods.

The spearman's correlation on the 1-month lag of the variables returned the results in Table 5 for each of the indicators for each of the TAMSAT and CHIRPS datasets.

Table 5: Spearman's Correlation of VCI3M against 1-month lag of TAMSAT/CHIRPS

| | RFE1M | RFE3M | RCI1M | RCI3M | SPI1M | SPI3M | Mean |
|---|---|---|---|---|---|---|---|
| TAMSAT | 0.23 | 0.39 | 0.33 | 0.64 | 0.38 | 0.64 | 0.44 |
| CHIPRS | 0.10 | 0.26 | 0.34 | 0.53 | 0.34 | 0.52 | 0.35 |

The TAMSAT variables are highly correlated to drought severity as compared to CHIRPS variables, except for RCI1M. Using bi-directional step-wise regression, the results offer a mixed case in which four variables were selected: TAMSAT_SPI3M_lag1, CHIRPS_RCI3M_lag1, CHIRPS_SPI3M_lag1 and TAMSAT_RCI1M_lag1. Though TAMSAT_SPI3M ranks high, CHIRPS variables are also as competitive. TAMSAT ranks, consistently higher on SPI that is widely used to drought monitoring. The Akaike information criterion (AIC), generally used to estimate the quality of a model relative to others had the 3-month aggregates of both SPI and RCI from TAMSAT as the best predictors. These results were confirmed by the relative importance of variables as partitioned by R^2 that also ranked SPI3M and RCI3M from TAMSAT ahead of the other variables. A final selection of variables using SVR and general additive model (GAM) techniques posted the result in Figure 6. The SVR model generally outperforms the GAM model for each of the variables except for TAMSAT_RCI3M and TAMSAT_SPI1M where similar performance is realized for the two models. The top performers in each case is are two TAMSAT variables.

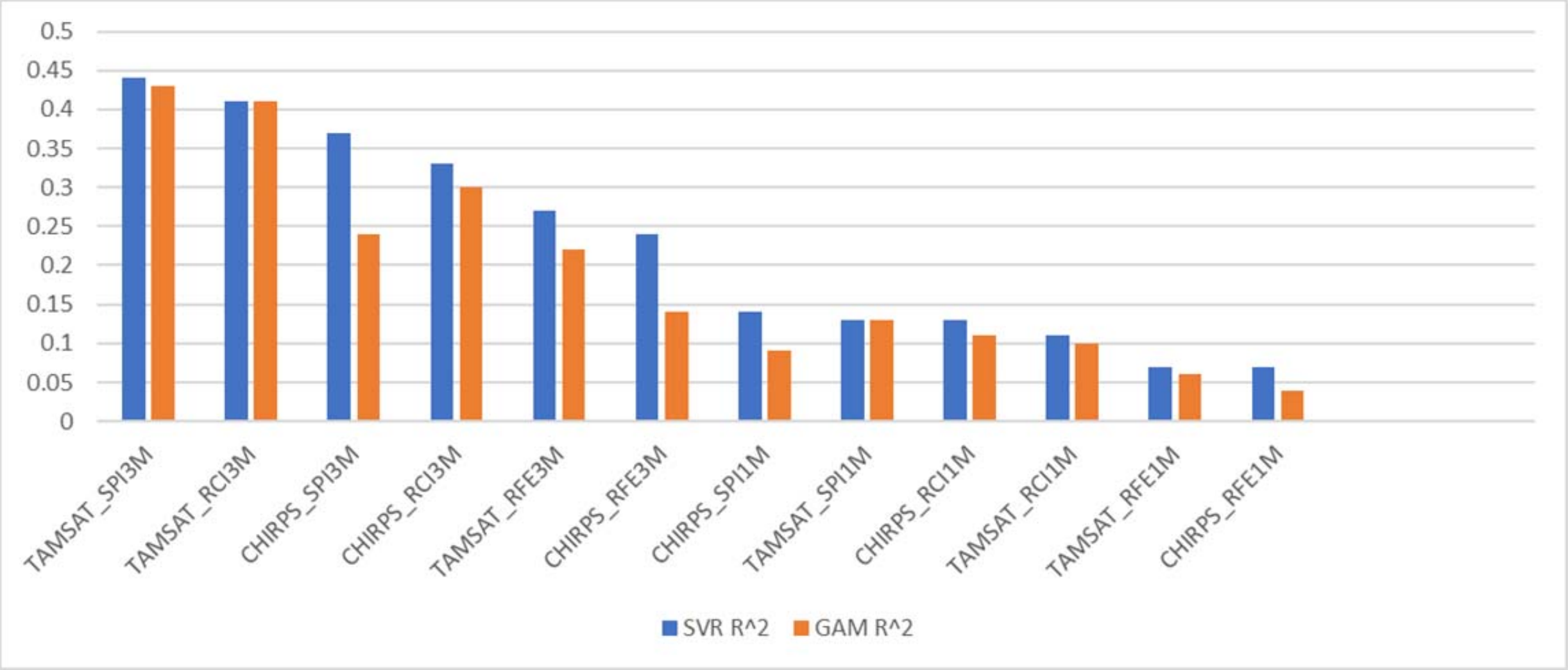

Figure 6: $R^2$ for SVR and GAM models for variable selection

From the multiple selection metrics, even though both precipitation datasets were competitive in drought prediction, TAMSAT generally produced better-ranked variables and was the dataset chosen for model building.

*3.2 Multi-collinearity of predictor variavles*

In investigation of possible multi-collinearity between the pairs of the predictor variables is provided in Figure 7.

| | NDVIDekad | VCI1M | VCIdekad | RCI1M | RCI3M | RFE1M | RFE3M | SPI1M | SPI3M | LST1M | EVT1M | PET1M | TCI1M | SPEI1M | SPEI3M |
|---|---|---|---|---|---|---|---|---|---|---|---|---|---|---|---|
| NDVIDekad | 1 | 0.54 | 0.54 | 0.16 | 0.32 | 0.18 | 0.49 | 0.03 | 0.32 | (0.51) | 0.65 | (0.37) | (0.20) | 0.12 | 0.25 |
| VCI1M | 0.54 | 1 | 1.00 | 0.18 | 0.52 | 0.07 | 0.29 | 0.19 | 0.54 | (0.32) | 0.51 | (0.22) | (0.41) | 0.08 | 0.25 |
| VCIdekad | 0.54 | 1.00 | 1 | 0.18 | 0.52 | 0.07 | 0.30 | 0.18 | 0.54 | (0.32) | 0.51 | (0.22) | (0.41) | 0.07 | 0.25 |
| RCI1M | 0.16 | 0.18 | 0.18 | 1 | 0.51 | 0.86 | 0.49 | 0.57 | 0.41 | (0.13) | 0.32 | (0.40) | (0.34) | 0.22 | 0.04 |
| RCI3M | 0.32 | 0.52 | 0.52 | 0.51 | 1 | 0.38 | 0.68 | 0.45 | 0.85 | (0.34) | 0.53 | (0.39) | (0.38) | 0.18 | 0.19 |
| RFE1M | 0.18 | 0.07 | 0.07 | 0.86 | 0.38 | 1 | 0.56 | 0.29 | 0.27 | (0.08) | 0.31 | (0.49) | (0.28) | 0.17 | 0.01 |
| RFE3M | 0.49 | 0.29 | 0.30 | 0.49 | 0.68 | 0.56 | 1 | 0.13 | 0.48 | (0.31) | 0.68 | (0.45) | (0.23) | 0.10 | 0.12 |
| SPI1M | 0.03 | 0.19 | 0.18 | 0.57 | 0.45 | 0.29 | 0.13 | 1 | 0.54 | (0.29) | 0.17 | (0.25) | (0.40) | 0.38 | 0.15 |
| SPI3M | 0.32 | 0.54 | 0.54 | 0.41 | 0.85 | 0.27 | 0.48 | 0.54 | 1 | (0.27) | 0.42 | (0.19) | (0.40) | 0.26 | 0.32 |
| LST1M | (0.51) | (0.32) | (0.32) | (0.13) | (0.34) | (0.08) | (0.31) | (0.29) | (0.27) | 1 | (0.54) | 0.68 | 0.48 | (0.29) | (0.26) |
| EVT1M | 0.65 | 0.51 | 0.51 | 0.32 | 0.53 | 0.31 | 0.68 | 0.17 | 0.42 | (0.54) | 1 | (0.43) | (0.31) | 0.09 | 0.12 |
| PET1M | (0.37) | (0.22) | (0.22) | (0.40) | (0.39) | (0.49) | (0.45) | (0.25) | (0.19) | 0.68 | (0.43) | 1 | 0.31 | (0.21) | (0.07) |
| TCI1M | (0.20) | (0.41) | (0.41) | (0.34) | (0.38) | (0.28) | (0.23) | (0.40) | (0.40) | 0.48 | (0.31) | 0.31 | 1 | (0.31) | (0.26) |
| SPEI1M | 0.12 | 0.08 | 0.07 | 0.22 | 0.18 | 0.17 | 0.10 | 0.38 | 0.26 | (0.29) | 0.09 | (0.21) | (0.31) | 1 | 0.78 |
| SPEI3M | 0.25 | 0.25 | 0.25 | 0.04 | 0.19 | 0.01 | 0.12 | 0.15 | 0.32 | (0.26) | 0.12 | (0.07) | (0.26) | 0.78 | 1 |

Figure 7: The collinearity matrix for X (predictors) variables

From the correlation matrix (Figure 7), the following are observed: (1) relatively high correlation (of up to 1 for VCI1M and NDVIDekad) between vegetation datasets. The assumption not to use multiple vegetation variables in the same model is thus justified. (2) SPI and RCI are highly correlated even though, in general, the pairings of precipitation data could be used in the same model. (3) the modifier variables of LST, EVT, PET, TCI and SPEI have acceptable correlation coefficients with vegetation and precipitation pairings. Grouping the variables in the modeling process into precipitation, vegetation and water balance (modifier) variables is thus a sound assumption.

*3.3 Correlation between predicted variable and all predictors*

The drought intensity formulated based on VCI3M and with higher values being essentially lower severity. As indicated in Table 6, the variables TCI, LST and PET are negatively correlated to VCI3M while all others have a positively correlation to VCI3M. With the assumption not to have two variables of the same type, it is clear that the problem of multi-collinearity is avoided given that the lags of variables highly correlated with VCI3M i.e. VCI3M, VCI1M and VCIdekad lagged by 1 month are not used together in the same model.

Table 6: Correlation between non-precipitation data and VCI3M

| Variable | Correlation with drought severity (VCI3M) |
|---|---|
| TCI1M_lag1 | -0.58 |
| LST1M_lag1 | -0.45 |
| PET1M_lag1 | -0.34 |
| NDVIDekad_lag1 | 0.16 |
| SPEI1M_lag1 | 0.19 |
| RFE1M | 0.23 |
| SPEI3M_lag1 | 0.28 |
| RCI1M | 0.33 |
| SPI1M | 0.38 |
| RFE3M | 0.39 |
| EVT1M_lag1 | 0.59 |
| RCI3M | 0.64 |
| SPI3M | 0.64 |
| VCI3M_lag1 | 0.82 |
| VCI1M_lag1 | 0.88 |
| VCIdekad_lag1 | 0.89 |

*3.4 Performance of ANN in model Training*

244 models from the different combination of the study variables were subjected to training using both ANN and SVR techniques to predict values of VCI3M 1 month ahead. For the ANN process, 15 models, marking 6.15% of the models, were found to have been overfit as indicated on the performance of the ANN models in Figure 8.

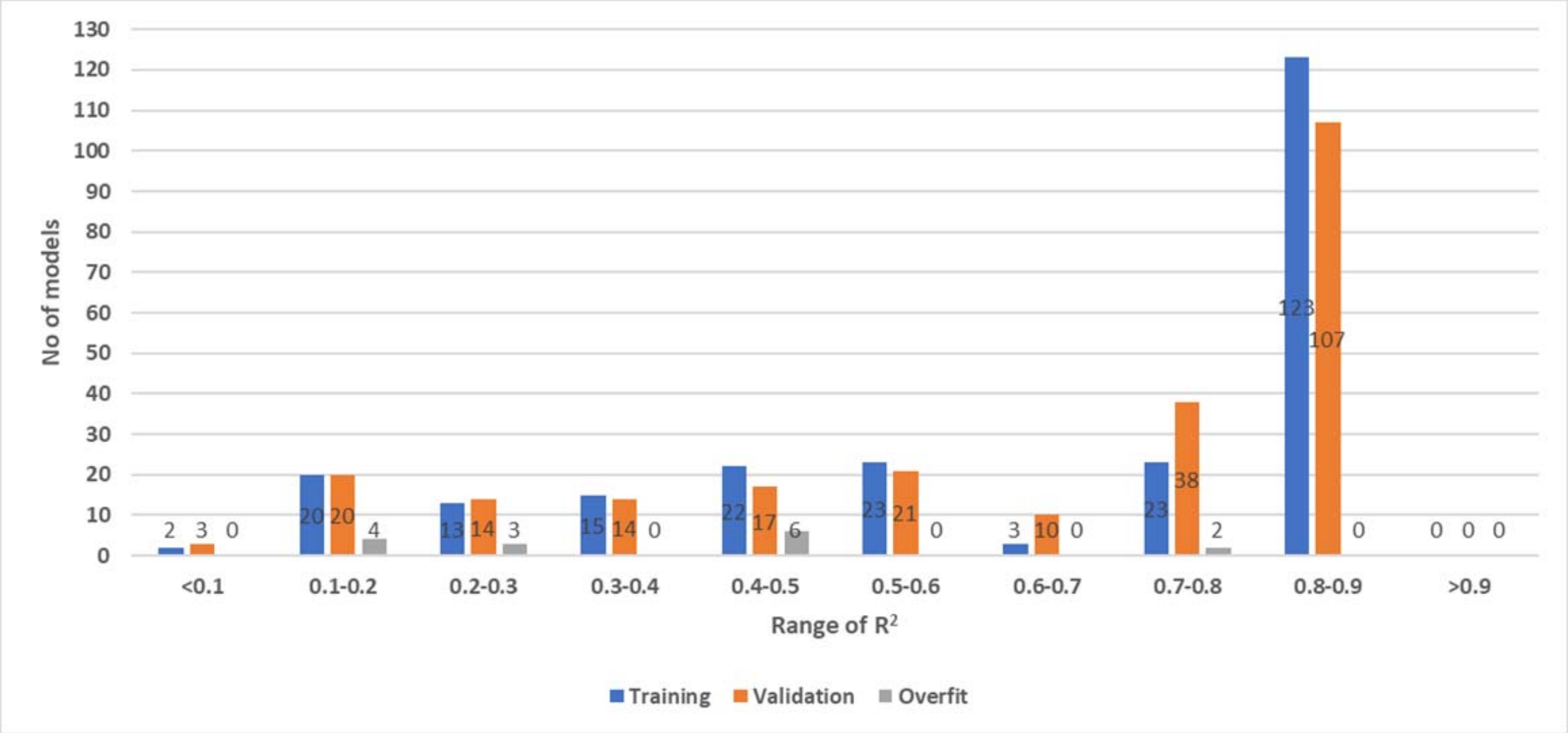

Figure 8: Performance of ANN models in the prediction of VCI3M grouped by descending performance ($R^2$)

From Figure 8, out of the 244 models subjected to the model training process, 145 models (59.43%) have an $R^2 \geq 0.7$ in the validation dataset. Interestingly, the overfitting is much less a problem amongst the models with $R^2 \geq 0.7$ as only 2 models representing 1.38% of the 145 models are considered to suffer over-fitting. The first 30 models ordered by decreasing $R^2$ in the validation dataset are presented in Table 7.

Table 7: Performance of the top 30 ANN models in training and validation ordered by descending $R^2$ in the validation dataset

| No | Model | $R^2$ Training | $R^2$ Validation | Overfit Index | Overfit |
|---|---|---|---|---|---|
| 1 | VCI3M_lag1 + TAMSAT_RCI3M_lag1 + SPEI1M_lag1 | 0.81 | 0.86 | 0.05 | 0 |
| 2 | VCIdekad_lag1 + TAMSAT_RFE1M_lag1 + TCI1M_lag1 | 0.87 | 0.86 | -0.01 | 0 |
| 3 | VCIdekad_lag1 + TAMSAT_RCI3M_lag1 + TCI1M_lag1 | 0.87 | 0.86 | -0.01 | 0 |
| 4 | VCIdekad_lag1 + TAMSAT_SPI3M_lag1 + TCI1M_lag1 | 0.87 | 0.86 | -0.01 | 0 |
| 5 | VCI1M_lag1 + TAMSAT_RCI3M_lag1 + TCI1M_lag1 | 0.87 | 0.86 | -0.01 | 0 |
| 6 | VCI3M_lag1 + TAMSAT_SPI3M_lag1 + SPEI1M_lag1 | 0.82 | 0.85 | 0.03 | 0 |
| 7 | VCI1M_lag1 + TAMSAT_SPI3M_lag1 + TCI1M_lag1 | 0.87 | 0.85 | -0.02 | 0 |
| 8 | VCIdekad_lag1 + TAMSAT_SPI1M_lag1 + PET1M_lag1 | 0.87 | 0.85 | -0.02 | 0 |
| 9 | VCIdekad_lag1 + TAMSAT_SPI1M_lag1 + TCI1M_lag1 | 0.86 | 0.85 | -0.01 | 0 |
| 10 | VCI1M_lag1 + TAMSAT_SPI3M_lag1 + PET1M_lag1 | 0.87 | 0.85 | -0.02 | 0 |
| 11 | VCIdekad_lag1 + TAMSAT_SPI3M_lag1 + LST1M_lag1 | 0.86 | 0.85 | -0.01 | 0 |
| 12 | VCIdekad_lag1 + TAMSAT_SPI3M_lag1 + PET1M_lag1 | 0.87 | 0.85 | -0.02 | 0 |
| 13 | VCIdekad_lag1 + TAMSAT_RFE3M_lag1 + TCI1M_lag1 | 0.87 | 0.85 | -0.02 | 0 |
| 14 | VCI1M_lag1 + TAMSAT_RFE1M_lag1 + TCI1M_lag1 | 0.87 | 0.85 | -0.02 | 0 |
| 15 | VCI1M_lag1 + TAMSAT_SPI3M_lag1 + LST1M_lag1 | 0.86 | 0.85 | -0.01 | 0 |
| 16 | VCIdekad_lag1 + TAMSAT_RCI3M_lag1 + PET1M_lag1 | 0.86 | 0.85 | -0.01 | 0 |
| 17 | VCI1M_lag1 + TAMSAT_RFE3M_lag1 + TCI1M_lag1 | 0.86 | 0.85 | -0.01 | 0 |
| 18 | VCIdekad_lag1 + TAMSAT_RCI3M_lag1 + SPEI1M_lag1 | 0.86 | 0.85 | -0.01 | 0 |
| 19 | VCI3M_lag1 + TAMSAT_RCI3M_lag1 + SPEI3M_lag1 | 0.8 | 0.84 | 0.04 | 0 |
| 20 | VCIdekad_lag1 + TAMSAT_SPI1M_lag1 + LST1M_lag1 | 0.86 | 0.84 | -0.02 | 0 |
| 21 | VCI1M_lag1 + TAMSAT_SPI1M_lag1 + PET1M_lag1 | 0.86 | 0.84 | -0.02 | 0 |
| 22 | VCI1M_lag1 + TAMSAT_SPI1M_lag1 + TCI1M_lag1 | 0.86 | 0.84 | -0.02 | 0 |
| 23 | VCIdekad_lag1 + TAMSAT_RCI3M_lag1 + LST1M_lag1 | 0.86 | 0.84 | -0.02 | 0 |
| 24 | VCI1M_lag1 + TAMSAT_RCI3M_lag1 + PET1M_lag1 | 0.86 | 0.84 | -0.02 | 0 |
| 25 | VCIdekad_lag1 + TAMSAT_RCI1M_lag1 + TCI1M_lag1 | 0.86 | 0.84 | -0.02 | 0 |
| 26 | VCIdekad_lag1 + TAMSAT_SPI3M_lag1 + SPEI1M_lag1 | 0.85 | 0.84 | -0.01 | 0 |
| 27 | VCIdekad_lag1 + TAMSAT_RFE1M_lag1 + LST1M_lag1 | 0.86 | 0.84 | -0.02 | 0 |
| 28 | VCI1M_lag1 + TAMSAT_RCI3M_lag1 + LST1M_lag1 | 0.86 | 0.84 | -0.02 | 0 |
| 29 | VCI1M_lag1 + TAMSAT_RFE1M_lag1 + LST1M_lag1 | 0.85 | 0.84 | -0.01 | 0 |
| 30 | VCIdekad_lag1 + TAMSAT_RFE1M_lag1 + SPEI1M_lag1 | 0.86 | 0.84 | -0.02 | 0 |

The top 30 models are noted to have no case of model over-fitting. In fact, 3 models out of the 30 models (10% of the top 30) are considered to be underfit models are therefore performed better in the validation dataset as compared to the training dataset. An extended analysis shows this trend as replicated in the top 100 models that have no occurrence of model over-fitting with 4% of the top 100 models actually underfit.

*3.5 Performance of SVR in model Training*

When subjected to the SVR process, 145 of the 244 models (59.43%) have $R^2 \geq 0.7$ in the validation dataset. This performance is comparable to the ANN technique. In terms of model overfitting, 21 of the 244 models (8.61%) are overfit as shown in Figure 9.

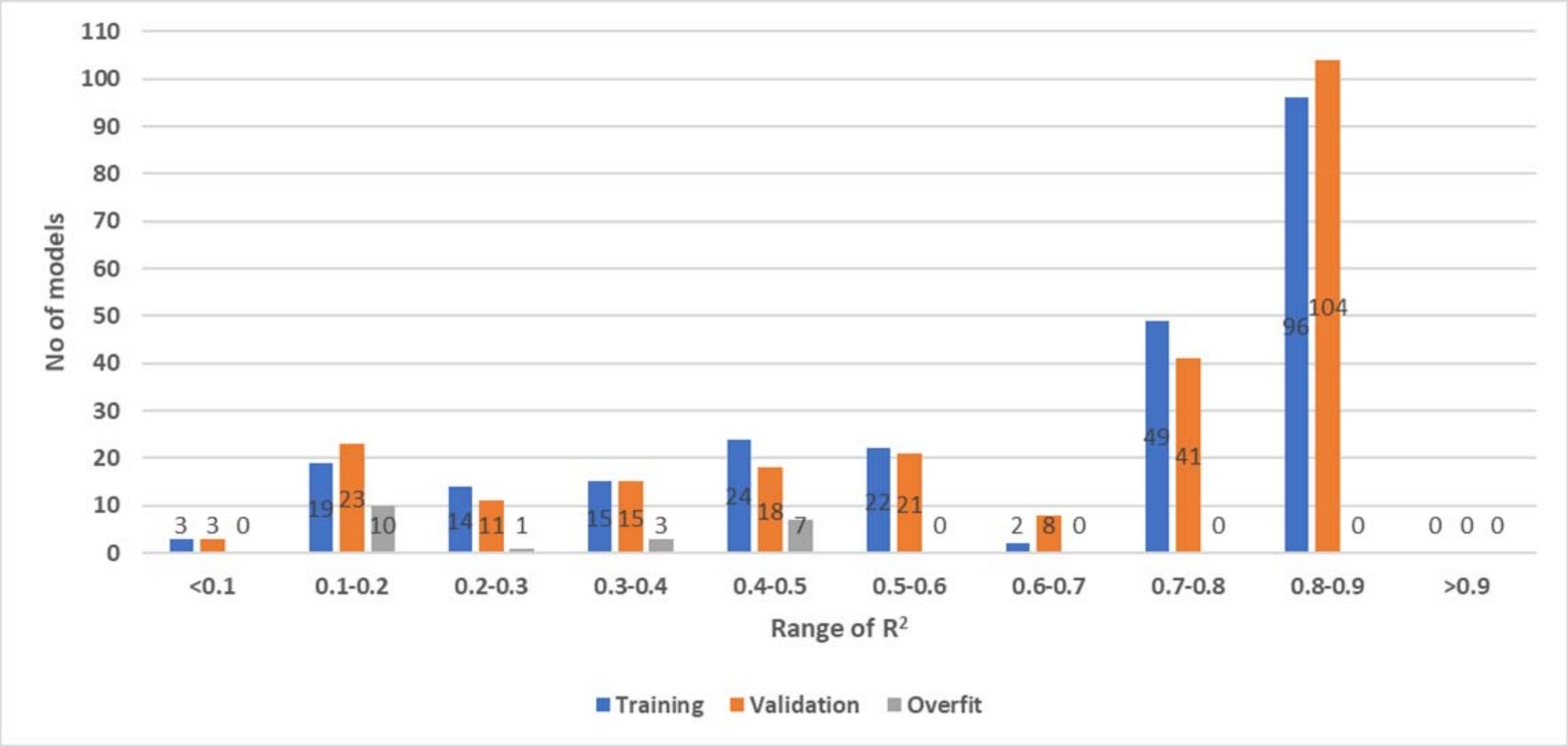

Figure 9: Performance of SVR models in the prediction of VCI3M by grouped by descending performance ($R^2$)

Like the case for ANN, there is no occurrence of overfitting in the top 30 and top 100 models ordered by descending $R^2$ in the validation dataset. In fact, the occurrence of overfitting for the SVR technique is confined to models with $R^2 \leq 0.5$. In as setting where selection of models is based on the $R^2 \geq 0.7$ cut off, the problem of model over-overfitting would be confined to the ANN technique as compared to the SVR technique. The tendency to suffer over-fitting in ANN with increase in performance is for example documented in Mitchel (1997). Like in the case for the ANN technique, we present, in Table 8, the performance in both training and validation the top 30 SVR models by descending $R^2$ in the validation dataset.

Table 8: Performance of the top 30 SVR models in training and validation ordered by descending $R^2$ in the validation dataset

| No | Model | $R^2$ Training | $R^2$ Validation | Overfit Index | Overfit |
|---|---|---|---|---|---|
| 1 | VCIdekad_lag1 + TAMSAT_SPI3M_lag1 + TCI1M_lag1 | 0.85 | 0.86 | 0.01 | 0 |
| 2 | VCIdekad_lag1 + TAMSAT_RCI3M_lag1 + TCI1M_lag1 | 0.85 | 0.86 | 0.01 | 0 |
| 3 | VCI3M_lag1 + TAMSAT_RCI3M_lag1 + SPEI1M_lag1 | 0.81 | 0.86 | 0.05 | 0 |
| 4 | VCI3M_lag1 + TAMSAT_SPI3M_lag1 + SPEI1M_lag1 | 0.82 | 0.85 | 0.03 | 0 |
| 5 | VCI1M_lag1 + TAMSAT_SPI3M_lag1 + TCI1M_lag1 | 0.85 | 0.85 | 0.00 | 0 |
| 6 | VCIdekad_lag1 + TAMSAT_SPI3M_lag1 + PET1M_lag1 | 0.84 | 0.85 | 0.01 | 0 |
| 7 | VCI1M_lag1 + TAMSAT_RCI3M_lag1 + TCI1M_lag1 | 0.85 | 0.85 | 0.00 | 0 |
| 8 | VCIdekad_lag1 + TAMSAT_SPI1M_lag1 + TCI1M_lag1 | 0.85 | 0.85 | 0.00 | 0 |
| 9 | VCIdekad_lag1 + TAMSAT_SPI3M_lag1 + LST1M_lag1 | 0.84 | 0.85 | 0.01 | 0 |
| 10 | VCIdekad_lag1 + TAMSAT_RFE1M_lag1 + TCI1M_lag1 | 0.85 | 0.85 | 0.00 | 0 |
| 11 | VCI1M_lag1 + TAMSAT_SPI3M_lag1 + PET1M_lag1 | 0.84 | 0.85 | 0.01 | 0 |
| 12 | VCI1M_lag1 + TAMSAT_SPI3M_lag1 + LST1M_lag1 | 0.84 | 0.84 | 0.00 | 0 |
| 13 | VCIdekad_lag1 + TAMSAT_SPI1M_lag1 + PET1M_lag1 | 0.84 | 0.84 | 0.00 | 0 |
| 14 | VCI1M_lag1 + TAMSAT_RFE1M_lag1 + TCI1M_lag1 | 0.85 | 0.84 | -0.01 | 0 |
| 15 | VCI1M_lag1 + TAMSAT_SPI1M_lag1 + TCI1M_lag1 | 0.85 | 0.84 | -0.01 | 0 |
| 16 | VCIdekad_lag1 + TAMSAT_RCI1M_lag1 + TCI1M_lag1 | 0.84 | 0.84 | 0.00 | 0 |
| 17 | VCIdekad_lag1 + TAMSAT_RCI3M_lag1 + LST1M_lag1 | 0.84 | 0.84 | 0.01 | 0 |
| 18 | VCIdekad_lag1 + TAMSAT_RFE3M_lag1 + TCI1M_lag1 | 0.84 | 0.84 | 0.00 | 0 |
| 19 | VCIdekad_lag1 + TAMSAT_RFE1M_lag1 + LST1M_lag1 | 0.84 | 0.84 | 0.00 | 0 |
| 20 | VCIdekad_lag1 + TAMSAT_SPI1M_lag1 + LST1M_lag1 | 0.84 | 0.84 | 0.00 | 0 |
| 21 | VCI3M_lag1 + TAMSAT_RCI3M_lag1 + TCI1M_lag1 | 0.83 | 0.84 | 0.01 | 0 |
| 22 | VCIdekad_lag1 + TAMSAT_RCI3M_lag1 + SPEI1M_lag1 | 0.83 | 0.84 | 0.01 | 0 |
| 23 | VCIdekad_lag1 + TAMSAT_SPI3M_lag1 + SPEI1M_lag1 | 0.84 | 0.84 | 0.00 | 0 |
| 24 | VCIdekad_lag1 + TAMSAT_RCI3M_lag1 + PET1M_lag1 | 0.83 | 0.84 | 0.01 | 0 |
| 25 | VCI1M_lag1 + TAMSAT_RCI3M_lag1 + LST1M_lag1 | 0.83 | 0.84 | 0.01 | 0 |
| 26 | VCI1M_lag1 + TAMSAT_SPI1M_lag1 + PET1M_lag1 | 0.84 | 0.84 | 0.00 | 0 |
| 27 | VCI3M_lag1 + TAMSAT_RCI3M_lag1 + SPEI3M_lag1 | 0.79 | 0.83 | 0.04 | 0 |
| 28 | VCI1M_lag1 + TAMSAT_RFE3M_lag1 + TCI1M_lag1 | 0.84 | 0.83 | -0.01 | 0 |
| 29 | VCIdekad_lag1 + TAMSAT_SPI3M_lag1 + EVT1M_lag1 | 0.83 | 0.83 | 0.00 | 0 |
| 30 | VCI1M_lag1 + TAMSAT_RCI1M_lag1 + TCI1M_lag1 | 0.84 | 0.83 | 0.00 | 0 |

*3.6 Comparative performance of the ANN & SVR techniques*

The analysis of the performance of the pairings of the 244 ANN and SVR models is as presented in Figure 10. The ANN and SVR techniques turned out competitive in model validation. 127 models (52%) posted similar performance with ANN outperforming SVR in 105 pairings (43%) and SVR outperforming ANN in 12 pairings (5%).

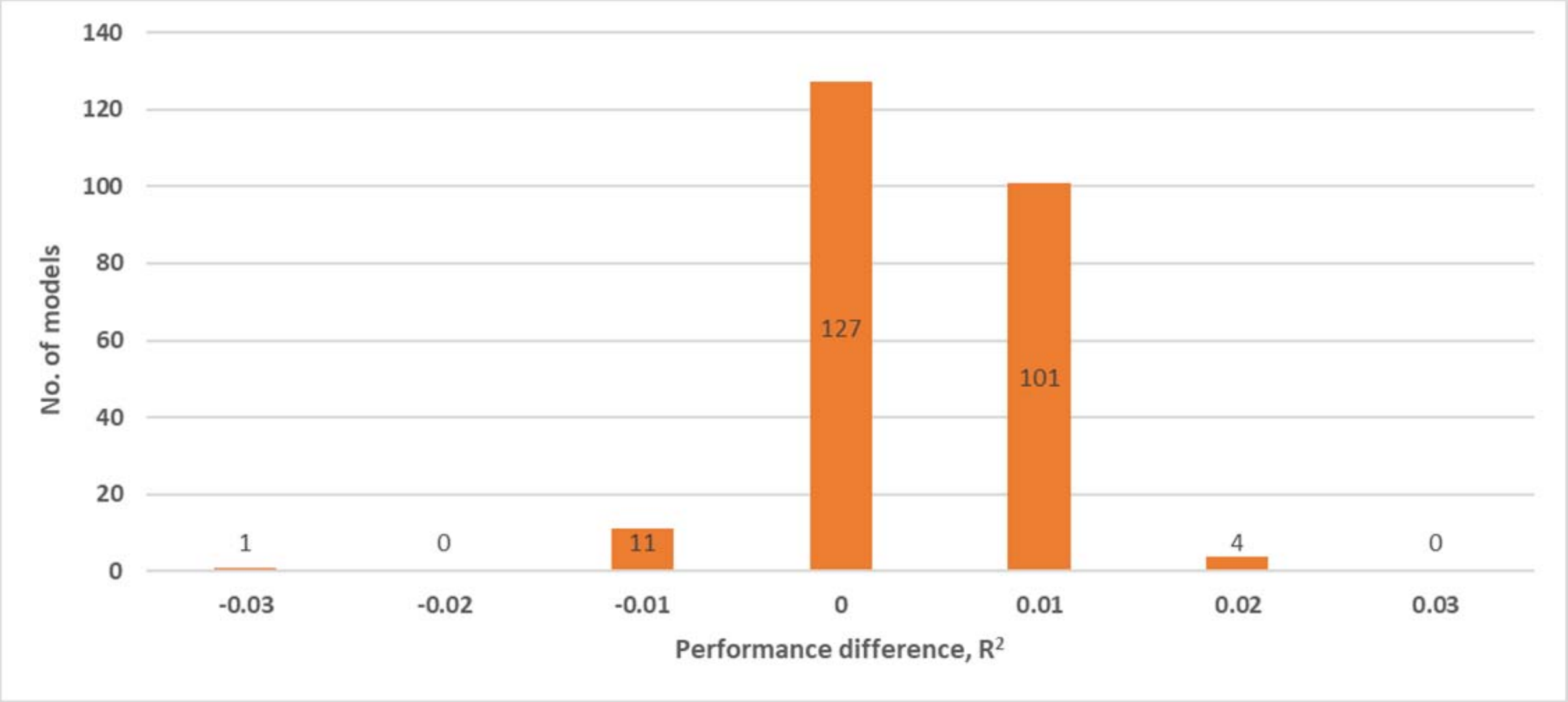

Figure 10: Performance difference between ANN and SVR model pairings.

The analysis of the competitiveness of the ANN and SVR techniques is done using the summary statistics of summary statistics of minimum, maximum, average on the 143 models that have $R^2 \geq 0.7$ and are not overfit from the ANN process. As indicated in Table 9, the techniques are quite competitive in this set of 143 models.

Table 9: Summary of performance ($R^2$) for each technique in model validation.

| Technique | Min | Max | Average | Range |
|---|---|---|---|---|
| SVR | 0.71 | 0.86 | 0.81 | 0.15 |
| ANN | 0.71 | 0.86 | 0.81 | 0.15 |

Given no models with $R^2 \geq 0.7$ were overfit from the SVR process, the choice of models for model ensembling was therefore a function of models from the ANN process. The selection of the appropriate models for ensemble membership was thus from the 143 ANN models paired with the corresponding SVR models of the same formula.

*3.3 Selection of ensemble membership*

Ensemble membership, also model pruning, is the selection phase of model ensembling. The different reasons for model selection as documented in Mendes-Moreira et al. (2012) are to reduce computational costs, if possible, to increase prediction accuracy and to avoid the multi-collinearity problem.

From the 143 models that had an $R^2 \geq 0.7$ and were not overfit, the construction of the ensemble membership faced two questions. First, was if all the 143 models were sufficient for the size of an ensemble and two, if there existed a smaller ensemble size that would perform the same, if not better than the 143. This question was answered following on the experimental process described in the methodology section and whose results are as plotted in Figure 11.

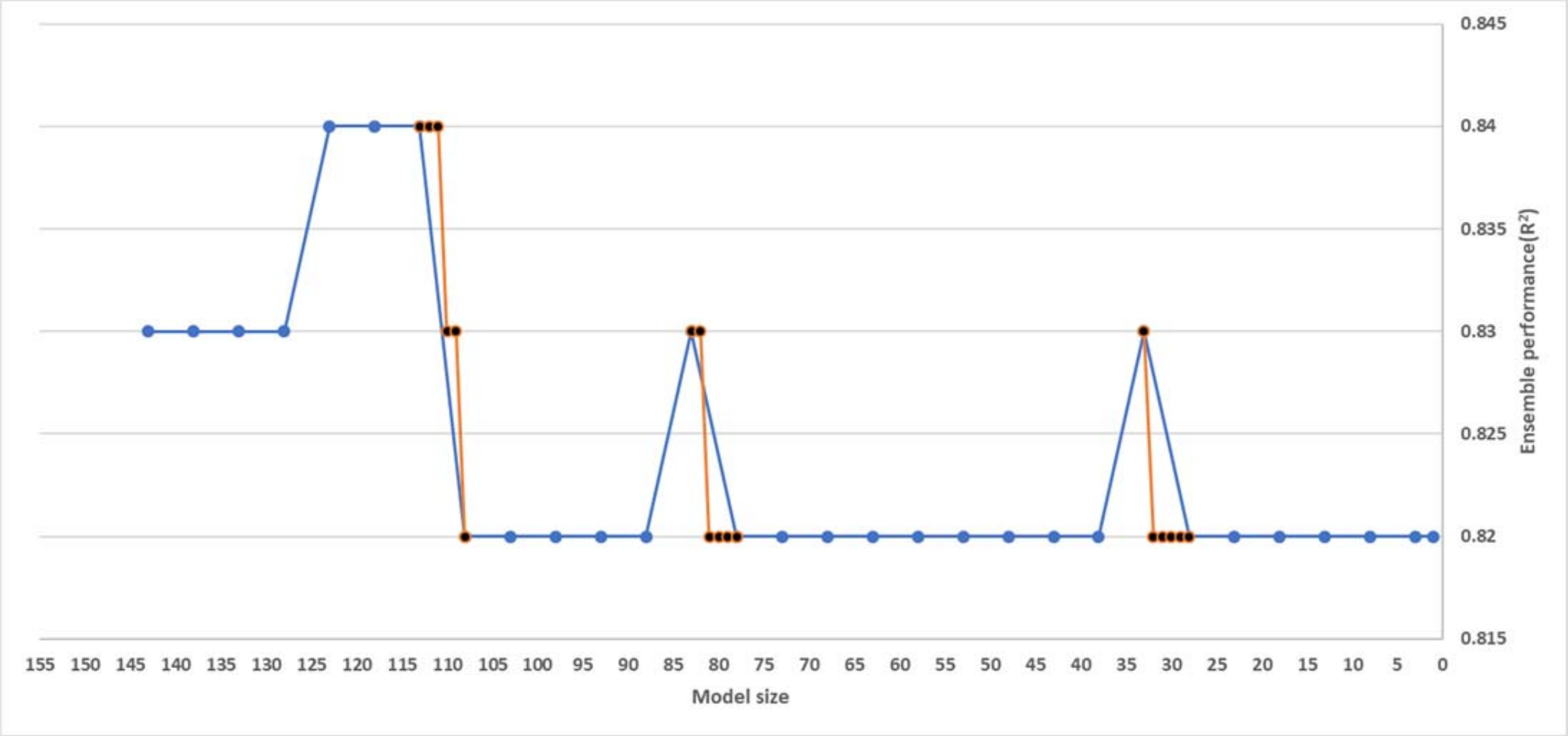

**Figure 11:** Ensemble membership selection showing the reduction from 143 models to 111 models in the ensemble using the back-forward selection procedure. The models are eliminated in batches of 5 but in instances where a drop of performance is realized, a forward selection beginning with the last smallest ensemble size is done by the addition of one model at a time (orange lines in the plot).

The elimination of the ranked models in batches of 5 and the forward selection in single units realizes the best smallest performance of the ensemble membership at a total of 111 models. For the ensemble membership, we opted for the size of 111 due to the trade-off of size and performance. This is guaranteed to have a reduced computational complexity associated with the ensemble size choice whilst not losing performance.

*3.6 Performance of ANN and SVR champion models in out-of-sample data*

In training, two different models are identified as the champion for each of the ANN and SVR techniques. Both the ANN and SVR champion models posted an $R^2$ of 0.86 in the validation dataset. The performance in training was also a comparable $R^2$ of 0.87 for the ANN to 0.85 for the SVR model.

The performance of the ANN and SVR processes in the out-of-sample dataset using the 96 data points is a contradiction of the performance in model training. The ANN champion posted a relatively stable performance of an $R^2$ of 0.82 (4% loss in performance) while the SVR champion loses performance to record an $R^2$ of 0.78. The above is despite the fact that a new SVR model arises with a higher performance than the SVR champion model. The performance in the test data at county level for both the ANN and the SVR is as provided in Table 10.

Table 10: Performance ($R^2$) of the champions models for each of the counties in the study area.

| | Mandera | Marsabit | Turkana | Wajir |
|---|---|---|---|---|
| ANN | 0.79 | 0.79 | 0.86 | 0.79 |
| SVR | 0.70 | 0.77 | 0.88 | 0.71 |

The results at county level indicates the consistency of ANN champion in out-performing the SVR champion except for Turkana county where SVR out performs ANN. Given that the models built are non-county specific, the performance of the best ANN and SVR models across the counties remains acceptable at an $R^2 \geq 0.79$ and $R^2 \geq 0.71$ for ANN and SVR respectively.

The study therefore uses the $R^2$ of from Table 9 as the baseline performance of the best model approach to model selection. These values will be used as the basis for comparison with the performance of both the homogenous and heterogenous model ensembles.

*3.6 Performance of homogenous model ensembles*

From the 111 models were selected for the investigation of model ensembles, we build both homogenous models of ANN and SVR independently. The performance of the homogenous ANN and SVR models were then investigated in both regression and classification.

3.6.1 Performance of homogenous ensembles in regression

The performance of the homogenous model ensembles in regression is provided in Table 11 and Table 12 for ANN and SVR techniques respectively. For each technique, results from the three ensembles approaches of non-weighted, weighted and stacked are presented.

Table 11: Performance ($R^2$) of the ANN homogenous model ensembles for each county. Each approach has the results derived from the non-weighted, weighted and stacked approaches to model ensembling.

| Approach | Mandera | Marsabit | Turkana | Wajir | Overall |
|---|---|---|---|---|---|
| ANN Champion | 0.79 | 0.79 | 0.86 | 0.79 | 0.82 |
| ANN Homogenous Simple Average | 0.78 | 0.86 | 0.88 | 0.80 | 0.84 |
| ANN Homogenous Weighted Average | 0.79 | 0.86 | 0.88 | 0.81 | 0.85 |
| ANN Homogenous Stacked | 0.93 | 0.87 | 0.89 | 0.93 | 0.91 |

Table 12: Performance ($R^2$) of the SVR homogenous model ensembles for each county. Each approach has the results derived from the non-weighted, weighted and stacked approaches to model ensembling.

| Approach | Mandera | Marsabit | Turkana | Wajir | Overall |
|---|---|---|---|---|---|
| SVR Champion | 0.70 | 0.77 | 0.88 | 0.71 | 0.78 |
| SVR Homogenous Simple Average | 0.71 | 0.80 | 0.87 | 0.73 | 0.80 |
| SVR Homogenous Weighted Average | 0.71 | 0.80 | 0.87 | 0.73 | 0.80 |
| SVR Homogenous Stacked | 0.88 | 0.85 | 0.88 | 0.88 | 0.88 |

For both ANN and SVR techniques, the model stacking approach to the formulation of model ensembles offers the best improvement as compared to the best model approach. In general, across the counties, the weighted together with the non-weighted approaches to model ensembling provide competitive performance to the best model approach.  posts mixed results when compared to the non-weighted approach. In fact, for both the weighted averaging and the simple averaging approach, the performance remains competitive as compared to the best model approach. Cases of loss of performance are also recorded using these two approaches like the case of SVR for Turkana county and simple average for Mandera county.

3.6.2 Performance of homogenous ensembles in classification

The summary of the performance of both the ANN and SVR homogenous ensembles is presented using both model accuracy in Table 13 and Table 14 respectively.

Table 13: Classification accuracy for the ANN homogenous ensembles.

| | Mandera | Marsabit | Turkana | Wajir | Overall |
|---|---|---|---|---|---|
| ANN Champion | 0.71 | 0.75 | 0.71 | 0.67 | 0.71 |
| ANN Homogenous Simple Average | 0.67 | 0.83 | 0.67 | 0.63 | 0.70 |
| ANN Homogenous Weighted Average | 0.67 | 0.79 | 0.67 | 0.63 | 0.69 |
| ANN Homogenous Stacked | 0.79 | 0.88 | 0.75 | 0.71 | 0.78 |

Table 14: Classification accuracy for the SVR homogenous ensembles.

| | Mandera | Marsabit | Turkana | Wajir | Overall |
|---|---|---|---|---|---|
| SVR Champion | 0.58 | 0.75 | 0.83 | 0.58 | 0.69 |
| ANN Homogenous Simple Average | 0.63 | 0.83 | 0.67 | 0.63 | 0.69 |
| ANN Homogenous Weighted Average | 0.63 | 0.83 | 0.71 | 0.63 | 0.70 |
| ANN Homogenous Stacked | 0.79 | 0.88 | 0.75 | 0.71 | 0.78 |

Using the three approaches to the building of model ensembles, for both the ANN and SVR approaches, it is clear that the homogenous ensembles are superior to the traditional champion model selection approach. The simple averaging (non-weighted) approach together with the weighted averaging approach are noted to offer improved performance for the homogenous ANN model ensembles. In the case of SVR homogenous model ensembles, even though performance gains are recorded as compared to the base champion model, it is the case that simple non-weighted averaging loses performance since there existed alternative models to the champion with higher performance.

*3.7 Performance of heterogenous model ensembles*

The performance of the heterogenous model ensembles of ANN and SVR was assessed both in regression and classification. Like in the case of homogenous models, we use the champion models (ANN & SVR) as the base models for the evaluation. Given that the predictions for the models were averaged for each model across the techniques, the still were the average of 222 models for each input data point in the test data.

### 3.7.1 Performance of heterogenous ensembles in regression

The performance of the heterogenous models is presented, at county level in Table 15 with the champion ANN and champion SVR as the base models.

Table 15: Performance ($R^2$) of the heterogenous model ensembles for each county. Each approach has the results derived from the non-weighted, weighted and stacked approaches to model ensembling.

| | **Mandera** | **Marsabit** | **Turkana** | **Wajir** | **Overall** |
|---|---|---|---|---|---|
| ANN Champion | 0.79 | 0.79 | 0.86 | 0.79 | 0.82 |
| SVR Champion | 0.70 | 0.77 | 0.88 | 0.71 | 0.78 |
| Heterogenous Simple Average | 0.74 | 0.82 | 0.87 | 0.76 | 0.82 |
| Heterogenous Weighted Average | 0.76 | 0.83 | 0.88 | 0.78 | 0.82 |
| Heterogenous Stacked | 0.94 | 0.94 | 0.91 | 0.96 | 0.94 |

With the champion model approach as the benchmark, loss of performance in regression is recorded for Mandera and Wajir counties for Simple Averaging and Weighted Averaging respectively with performance in Turkana considered relatively stable for these two approaches. The heterogenous stacked approach in regression is seen to offer the best improvement in performance across each of the counties and on the entire dataset. The improvement in performance by $R^2$ using the heterogenous stacked approach and the ANN champion as the base ranges from 0.05 (Turkana) to 0.17 in Wajir.

### 3.7.2 Performance of heterogenous ensemble in classification

The classification accuracy of the heterogenous ensemble as compared to that of the champion models is presented in Table 16.

Table 16: Classification accuracy of the heterogenous ensemble

| | **Mandera** | **Marsabit** | **Turkana** | **Wajir** | **Overall** |
|---|---|---|---|---|---|
| ANN Champion | 71 | 75 | 71 | 67 | 71 |
| SVR Champion | 58 | 75 | 83 | 58 | 69 |
| Heterogenous Simple Average | 63 | 83 | 71 | 63 | 70 |
| Heterogenous Weighted Average | 63 | 83 | 71 | 67 | 71 |
| Heterogenous Stacked | 71 | 88 | 79 | 83 | 80 |

The accuracy of the heterogenous model ensembles in classification (Table 15), shows quite an overall improvement in performance of 9 and 11% for the ANN and SVR processes respectively. For Turkana county, the SVR champion however outperforms the ANN champion and the three approaches to homogenous model ensembling.

### 3.7.3 Further analysis of the performance of the heterogenous stacked ensembles

Given the superiority of the stacking approach to model ensembling as applied to heterogenous ensembles, we present a further analysis of their performance- both in regression and in classification of the out-of-sample data.

The performance of the heterogenous stacked approach to ensembling in regression is presented in Figure 12 for each of the 24 data points in the out-of-sample datasets of the counties in the study area.

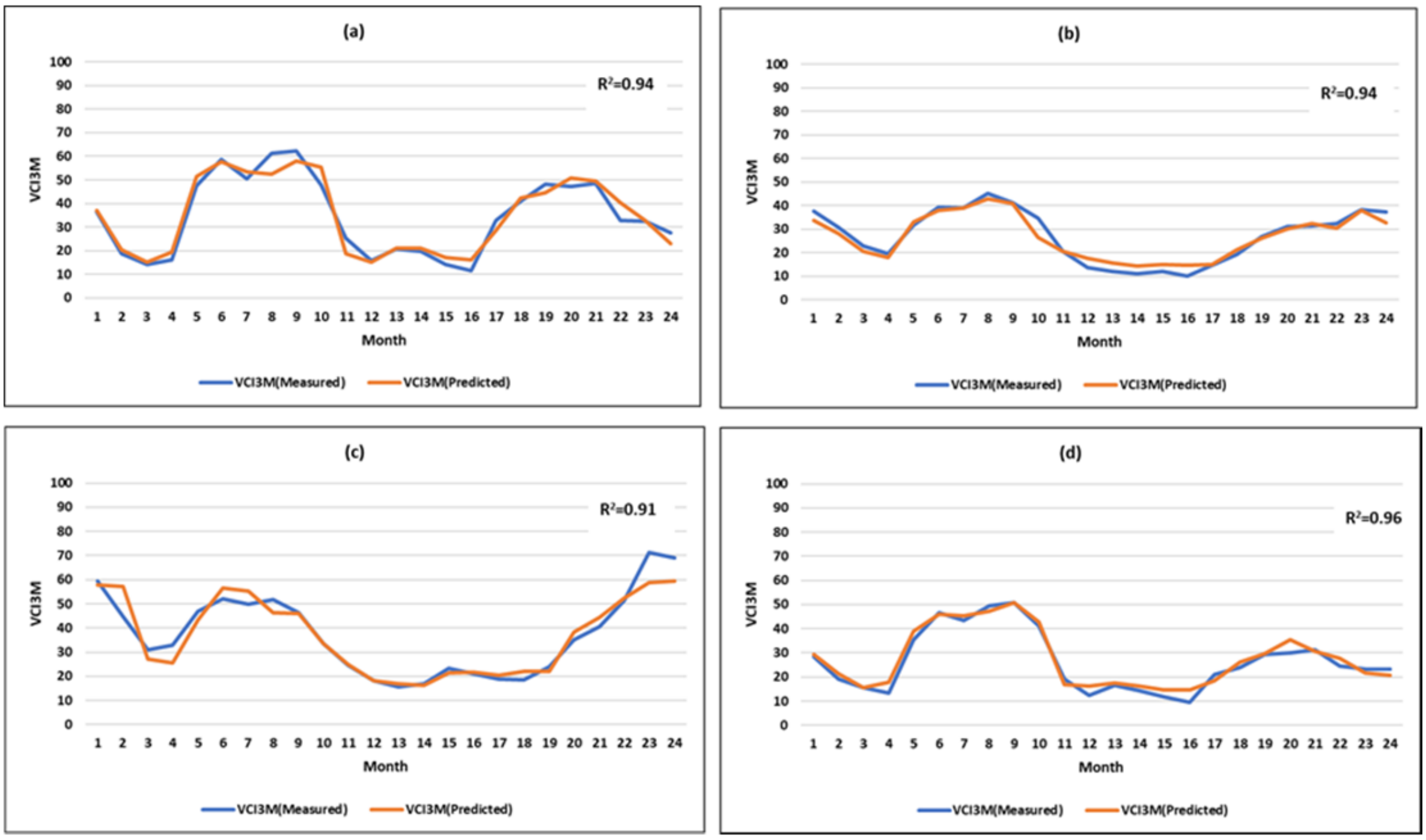

**Figure 12.** Plot of the actual values of VCI3M versus the values predicted 1 month ahead from the heterogenous stacked ensembles in the test data over 24 months for (a) Mandera ($R^2$=0.94); (b) Marsabit ($R^2$=0.94); (c) Turkana ($R^2$=0.91) and (d) Wajir ($R^2$=0.96).

The heterogenous stacked model, as indicated in Figure 12 posts quite a good agreement between the measured vegetation conditions as compared to the predicted values with an $R^2$ of between 0.91 and 0.96 for the counties. In classification, the stacking approach on the heterogenous models realizes the performance presented earlier in Table 16. The month by month performance of the heterogenous stacked classifier is as presented in Figure 13.

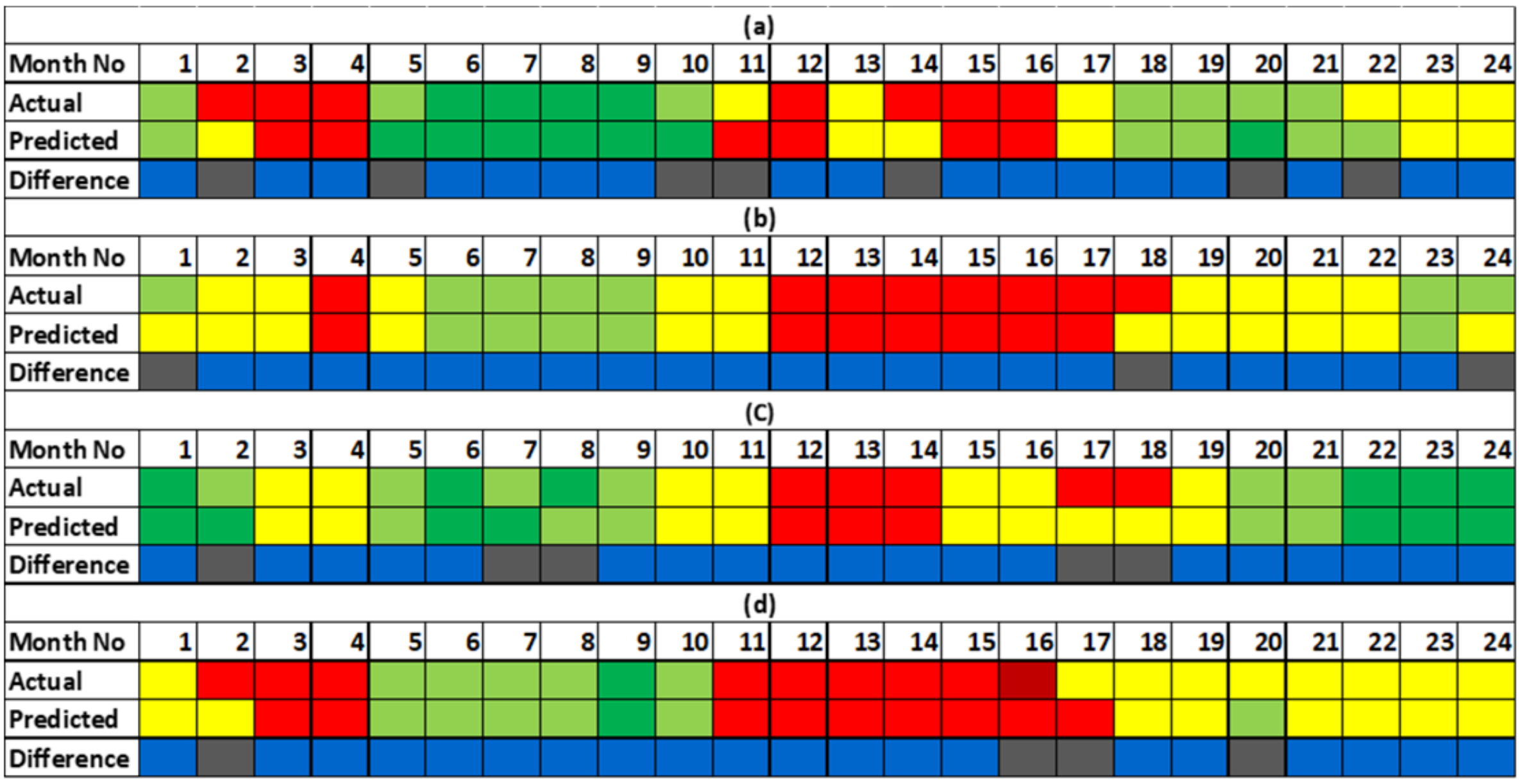

**Figure 13.** Performance of the heterogenous ensemble classifier for the each of the counties showing months of difference in grey and those of agreement in blue. Predictions are done 1 month ahead. The classification accuracies are: **(a)** 71% for Mandera county; **(b)** 88% for Marsabit county; **(c)** 79% for Turkana county and; **(d)** 83% for Wajir county.

The months in blue indicate when the class prediction is correct while grey indicates non-correct predictions. This illustrates the heterogenous stacked ensemble accuracy range from a best of 88% in Marsabit to a minimum of 71% for Mandera. Clearly, the performance of the heterogenous stacked classifiers with an overall accuracy of 80% is superior to that of the best champion model (ANN) that posts accuracy of 71% over the entire test dataset. Even any county level, the heterogenous stacked classifier outperforms the champion classifiers across all counties except for Turkana where the SVR champion performs better.

Moderate to extreme vegetation deficits correspond to the occurrence of drought events. Their correct prediction forms the best judgment on the utility of the outputs of the modelling approaches. Table 16 presents the performance in the prediction of moderate to extreme vegetation deficits across the different classes for the counties in the study area.

Table 16: Performance in the prediction of moderate to extreme drought of the heterogenous stacked ensembles compared to the best ANN and SVR models

| County | ANN Champion | SVR Champion | Heterogenous Stacked Ensemble |
|---|---|---|---|
| Mandera | 0.62 | 0.46 | 0.69 |
| Marsabit | 0.71 | 0.71 | 0.94 |
| Turkana | 0.75 | 1.00 | 0.83 |
| Wajir | 0.72 | 0.61 | 0.78 |
| Overall | 0.70 | 0.69 | 0.82 |

The heterogenous stacked ensemble is shown in Table 16 to offer the best performance in the prediction of moderate to extreme vegetation deficit. The distributions across the counties is also acceptable as compared, for example, to the SVR champion that performs below chance (accuracy less than 50%) for Mandera county.

**4. Conclusion**

The precipitation datasets from both TAMSAT and CHIRPS were not normally distributed. A multiple metrics investigation indicated the marginal superiority of TAMSAT and compared to CHIRPS in correlation to drought intensity. The use of the backward-forward approach in the selection of models for ensemble membership was realized to be a viable one as it reduced the model space to a total of 111 models from a set of 143 models.

The traditional approach to model selection that ends up with one champion model based on their performance on the validation dataset is shown to be prone to loss of model performance as evidenced by the SVR process in which a loss in performance from 0.83 to 0.78 was recorded. The building of model ensembles would not only guarantee stability but ensure increased accuracy in model performance.

The model ensembling approaches investigated in the study included non-weighted averaging, weighted averaging and model stacking as applied to both homogenous and heterogenous model ensembling approaches. Empirically, it was shown that heterogenous ensembles are generally more robust as compared to homogenous ensembles. Also, model stacking is indicated to be the surest way to realize model ensembles that are better in performance as compared to the champion model approach. In-fact, it is empirically shown that a loss in performance could be suffered when the averaging approaches are used especially when the models in the ensembles are selected based on a common performance metric.

The performance of the models learnt using the heterogenous model stacking approach are noted to be robust both in terms of regression and classification and also in the performance when generalized for the individual units even when models learn were not administrative unit specific.

This is a key finding since it implies the approach is robust enough to learn a single model applicable in the prediction across multiple administration units.

The performance of the models in the prediction of moderate to extreme vegetation deficit for the models records an $R^2$ of between 0.69 and 0.94 at county level. Given that the prediction of these conditions forms the practical application for the ensemble models for a good performance and a guaranteed utility of the forecast since they are way better than chance.

The study however, advices the use of more techniques in the model ensembles and the building of many more ensembles using different ensemble sizes to fully settle the question of performance of model ensembles.

There is a guaranteed gain in performance in using model stacking as an approach to building homogenous model ensembles. This is evidenced by the 4-13% improvement in accuracy for the ANN homogenous ensemble across the four counties.

**Author Contributions:** Conceptualization, Chrisgone Adede; Formal analysis, Chrisgone Adede; Investigation, Chrisgone Adede; Methodology, Chrisgone Adede, Robert Oboko, Peter Wagacha and Clement Atzberger; Supervision, Robert Oboko and Peter Wagacha; Validation, Robert Oboko, Peter Wagacha and Clement Atzberger; Visualization, Chrisgone Adede; Writing – original draft, Chrisgone Adede, Robert Oboko and Peter Wagacha; Writing – review & editing, Clement Atzberger.

**Funding:** This research received no direct external funding. The data used in the study was however, partly funded by the European Commission's funding under a grant contact to the Institute for Surveying, Remote Sensing and Land Information, University of Natural Resources and Life Sciences (BOKU).

**Acknowledgements:** Our appreciation to the National Drought Management Authority for providing the data from the operational drought monitoring system. We are also grateful to Luigi Luminari for the continued discussion of the ideas of the paper towards shaping it to have outputs applicable in an operational drought monitoring environment. The helpful contribution of the editors and reviewers are also acknowledged.

**Conflict of Interest:** The authors declare no conflict of interest.